\documentclass[journal]{IEEEtran}

\ifCLASSINFOpdf
\usepackage[pdftex]{graphicx}
\graphicspath{{../pdf/}{../jpeg/}}
\DeclareGraphicsExtensions{.pdf,.jpeg,.png}
\else
\usepackage[dvips]{graphicx}
\graphicspath{{../eps/}}
\DeclareGraphicsExtensions{.eps}
\fi
\usepackage{cite}
\usepackage{algorithmic}
\usepackage{amssymb}
\usepackage{textcomp}

\usepackage{amsmath}
\usepackage{amsfonts}
\usepackage{flushend}
\usepackage{amssymb}
\usepackage{makecell}
\usepackage{booktabs}

\ifCLASSOPTIONcompsoc

\usepackage[caption=false,font=normalsize,labelfont=sf,textfont=sf]{subfig}
\else
\usepackage[caption=false,font=footnotesize]{subfig}
\fi
\usepackage{graphicx}

\usepackage{epsfig}
\usepackage{epstopdf}
\usepackage{color}

\let\emph\textit
\usepackage{fixltx2e}

\usepackage{multicol}  
\usepackage{multirow}

\def\BibTeX{{\rm B\kern-.05em{\sc i\kern-.025em b}\kern-.08em
		T\kern-.1667em\lower.7ex\hbox{E}\kern-.125emX}}

\begin{document}
	
\title{Empirical Studies of Propagation Characteristics and Modeling Based on XL-MIMO Channel Measurement: From Far-Field to Near-Field }

%
%
%

\author{Haiyang~Miao, \IEEEmembership{Graduate Student Member,~IEEE}, Jianhua~Zhang, \IEEEmembership{Senior Member,~IEEE}, Pan~Tang, \IEEEmembership{Member,~IEEE}, Lei Tian, \IEEEmembership{Member,~IEEE}, Weirang Zuo, Qi Wei, Guangyi Liu, \IEEEmembership{Member,~IEEE}

	\thanks{This work was supported by the National Science Fund for Distinguished Young Scholars under Grant 61925102, National Natural Science Foundation of China (62101069 \& 62201086 \& 92167202), and Beijing University of Posts and Telecommunications-China Mobile Research Institute Joint Innovation Center. (\itshape{Corresponding author: Jianhua Zhang and Pan Tang.})}
	\thanks{H. Miao, J. Zhang, P. Tang, L. Tian, W. Zuo, and Q. Wei are with State Key Laboratory of Networking and Switching Technology, Beijing University of Posts and Telecommunications, Beijing 100876, China (e-mail: hymiao@bupt.edu.cn; jhzhang@bupt.edu.cn; tangpan27@bupt.edu.cn; tianlbupt@bupt.edu.cn; zuoweirang@bupt.edu.cn; weiqi0814@bupt.edu.cn).}
	\thanks{G. Liu is with Future Research Laboratory, China Mobile Research Institute, Beijing 100053, China (e-mail: liuguangyi@chinamobile.com).}
    }


\markboth{Journal of \LaTeX\ Class Files,~Vol.~14, No.~8, August~2021}%
{Shell \MakeLowercase{\textit{et al.}}: A Sample Article Using IEEEtran.cls for IEEE Journals}

%

\maketitle

\begin{abstract}
	
	 In the sixth-generation (6G), the extremely large-scale multiple-input-multiple-output (XL-MIMO) is considered a promising enabling technology. With the further expansion of array element number and frequency bands, near-field effects will be more likely to occur in 6G communication systems. The near-field radio communications (NFRC) will become crucial in 6G communication systems. It is known that the channel research is very important for the development and performance evaluation of the communication systems. In this paper, we will systematically investigate the channel measurements and modeling for the emerging NFRC. First, the principle design of massive MIMO channel measurement platform are solved. Second, an indoor XL-MIMO channel measurement campaign with 1600 array elements is conducted, and the channel characteristics are extracted and validated in the near-field region. Then, the outdoor XL-MIMO channel measurement campaign with 320 array elements is conducted, and the channel characteristics are extracted and modeled from near-field to far-field (NF-FF) region. The spatial non-stationary characteristics of angular spread at the transmitting end are more important in modeling. We hope that this work will give some reference to the near-field and far-field research for 6G.
	
\end{abstract}


\IEEEpeerreviewmaketitle

\section{Introduction}

\subsection{Background}

\par At present, the researches on sixth-generation (6G) has been carried out on a global scale [1-4]. In June 2023, the 44th meeting of International Telecommunication Union-Radiocommunication Sector (ITU-R) WP 5D described the 6G overall objectives and trends [5]. For fulfilling these goals and trends, tremendous research efforts are being made to develop new wireless technologies to meet the key performance indicators of 6G, which are superior to those of fifth-generation (5G). For instance, thanks to the enormous spatial multiplexing and beamforming gain, the extreme multiple-input-multiple-output (E-MIMO) is expected to accomplish a 10-fold increase in the spectral efficiency for 6G [1]. E-MIMO would be deployed with new types of antenna arrays, such as extremely large-scale MIMO (XL-MIMO), a distributed mechanism and so on [5-6]. This is expected to achieve better spectrum efficiency, larger network coverage, more accurate positioning, higher energy efficiency, etc [7-9]. Meanwhile, the size of antenna array is further increased in E-MIMO communication systems, which will bring unique channel characteristics and need to explore new channel models. In December 2023, the 3GPP Rel-19 has determined the need to study the near-field channel model for the mid-band. A widely used metric for determining the boundary between far-field and near-field region is the Rayleigh distance $2D^2/\lambda$, also known as the Fraunhofer distance [10]. Fig. 1 shows that the near-field MIMO communications will become essential in 6G networks. The $D$ and $f$ are the aperture of antenna array and carrier frequency respectively.

\begin{figure}[htbp]
	\setlength{\abovecaptionskip}{0.1 cm}
	\centering
	\includegraphics[width=0.42\textwidth]{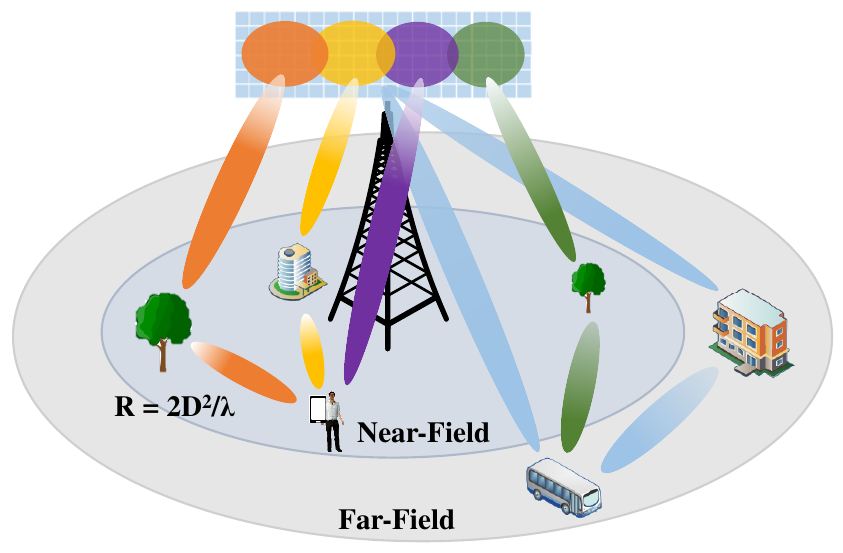}
	\caption{Near-field and far-field in the wireless communication.}
	\label{fig:Figure1}
\end{figure}

\subsection{Related Progress}

\par At present, some research has been carried out to explore the channel for large scale antenna array. In [11-12], the channel measurements were performed at 2.6 GHz band, with a base station (BS) equipped with the virtual uniform linear array (ULA) and uniform cylindrical array (UCA) of 128 ports. The investigation showed that the measured channels can achieve performance close to that in i.i.d. Rayleigh channels for both array types. In [13], the received power and average power delay distribution (APDP) were studied in 2.6 GHz band, and the non-stationarity was analyzed in the spatial domain. In [14-15], two array configurations (including 128-element virtual ULA and UCA) were used to conduct channel measurement in an outdoor stadium scenario in the 1.4725 GHz band, and the angular power spectrum along the virtual linear array was analyzed. In [16], to gain further insight into the massive MIMO channel, the channel information was extracted from measurement data in the urban scenario. In [17], the delay spread characteristic of a massive MIMO system was analyzed at 3.5 GHz in outdoor-to-indoor (O2I) scenario. In [18], the virtual 16$\times$60 antenna UCA at Tx and the dual-polarized 64-port uniform circular patch array at Rx were used for channel measurement in the urban scenario in 2.53 GHz band. In [19], the statistical properties of massive MIMO channels were studied at low band. In [20], the indoor mmWave massive MIMO channel measurements of 64 and 128 elements were studied in 26 GHz band. In [21], the multi-frequency MIMO channel characteristics were studied at mmWave bands. In [22], an indoor channel measurement was presented from 26 to 30 GHz, using the massive virtual array of 21$\times$21 elements. In [23], a terahertz (THz) ultra-massive MIMO architecture was introduced and the model-based and model-free hybrid beamforming techniques were investigated.

\par In [24], the evolution of cluster number with 256 virtual array was given in the 3.5 GHz band under line-of-sight (LOS) and non-line-of-sight (NLOS) conditions respectively, which showed the cluster appearance and disappearance, and established the corresponding model based on the birth-death process. In [25], we proposed a channel modeling method based on the capture and observation of multipath propagation mechanisms at mmWave band. In [26], two neural network (NN) models were developed for modeling of path loss together with shadow fading (SF) and small-scale channel parameters. In [27], to enable realistic studies of massive MIMO systems, the COST 2100 channel model was extended based on measurements. The COST channel model defines the physical location of the scatterer, not directly angles of departure or arrival as seen from the terminal, which is difficult to extract the COST model parameters from measurements. Most importantly, these channel modeling methods differ greatly from the 5G channel modeling ideas standardized by 3GPP, and cannot continue to evolve based on the 5G standard model [28], which is also an important challenge encountered in the current 6G channel standardization.

\subsection{Motivation}

\par Most channel measurements were carried out at 1.4725 GHz [15-16], 2.53 GHz [19], 2.6 GHz [12-14], 3.5 GHz [17-18], and high-bands [20-24]. Some existing work has observed near-field spatial nonstationarity, but there is no characterization and modeling analysis of nonstationarity on array or spatial distance. However, this is very important for the quantitative analysis of near-field characteristics, such as the parameter setting in the 3GPP simulation modeling process, which is also concerned in 3GPP Rel-19.

\par Currently, the new mid-band (6-24 GHz) has attracted extensive attention from academia and industry, becoming the potential band for 6G. At World Radiocommunication Conference 2019 (WRC-19), it was decided that the 6 GHz band will be included to the WRC-23 agenda item 1.2 as the new IMT (5G or 6G) band. The 3GPP RAN Plenum adopted a standard modification proposal for the band of 5925-7125 MHz, which is introduced into 5G-Advanced Rel-18. In addition, some companies and researchers expressed their considerations for 6G spectrum. Samsung considers an in-depth study is needed on how to use the 6-20 GHz for 6G [29]. China Mobile holds the considerations on the 6 GHz spectrum for 5G-Advanced and 6G [30]. In [31], the system performances at the new mid-band were discussed by comparing with sub-6 GHz and mmWave bands. Notably, in November 2023, the WRC-23 proposed to identify a new band (6 GHz band) for mobile communication in every ITU Region [32].

\subsection{Main Contributions}

\par In this paper, we present the channel measurement, theoretical derivation, and modeling from near-field to far-field for XL-MIMO. The main contributions and novelties of this paper are summarized as follows:

\begin{itemize}

\item The massive MIMO broadband channel measurement platform covering mid-band is built and validated, which is based on the high-speed electronic switching matrix with nanosecond switching accuracy. Based on the platform, the XL-MIMO channel measurements are carried out in the near-field and far-field region.

\item The new channel characteristics are validated based on the indoor near-field channel measurements. Especially, the near-field channel characteristics on antenna array are validated by investigating the temporal-spatial channel characteristics including power, delay spread, angular spread, etc.

\item The channel characteristics are investigated and modeled from the near-field to far-field (NF-FF). The spatial non-stationarity considering antenna array and transmit-receive distance are investigated and modeled the temporal-spatial channel characteristics including path loss, delay spread, and angular spread. Compared with the receiving end, the spatial non-stationary characteristics of angular spread at the transmitting end are more obvious.

\end{itemize}

\par The rest of this paper is organized as follows. In Section II, the design and construction of massive MIMO channel measurement platform are discussed. In Section III, based on the XL-MIMO channel measurements, channel characteristics are validated and modeled from near-field to far-field. In Section IV, the near-field channel model is deduced theoretically, and the XL-MIMO channel model simulation framework is evolved and validated. Finally, we conclude this paper in Section V.


\section{Channel Measurement Platform Design and Data Post-Processing}

\par It is known that due to the limitations of hardware, technology and other aspects, the virtual measurement method with a single antenna is often used in massive MIMO channel measurement, which limits the effective capture of spatial characteristics. This section will focus on the MIMO channel measurement platform based on an electronically switched antenna array, which can be dynamically adjusted according to the signal symbol rate. Therefore, the issue needs to be solved, i.e., How to design the platform based on theoretical principles?

\subsection{Measurement Platform}

\par Fig. 2 shows the massive MIMO channel measurement platform based on high-speed electronic switching switch, making it a reality to measure the comprehensive characteristics of mid-band MIMO channel based on TDM-MIMO mode. Combined with pseudo-random (PN) sequence signal and TDM antenna switching technology, it can decouple the response of the antenna measurement system, to understand the transmission channel with high precision, and effectively solve the measurement limitations of mid-band channel.


\begin{figure}[htbp]
	\setlength{\abovecaptionskip}{0.1 cm}
	\centering
	\includegraphics[width=0.45\textwidth]{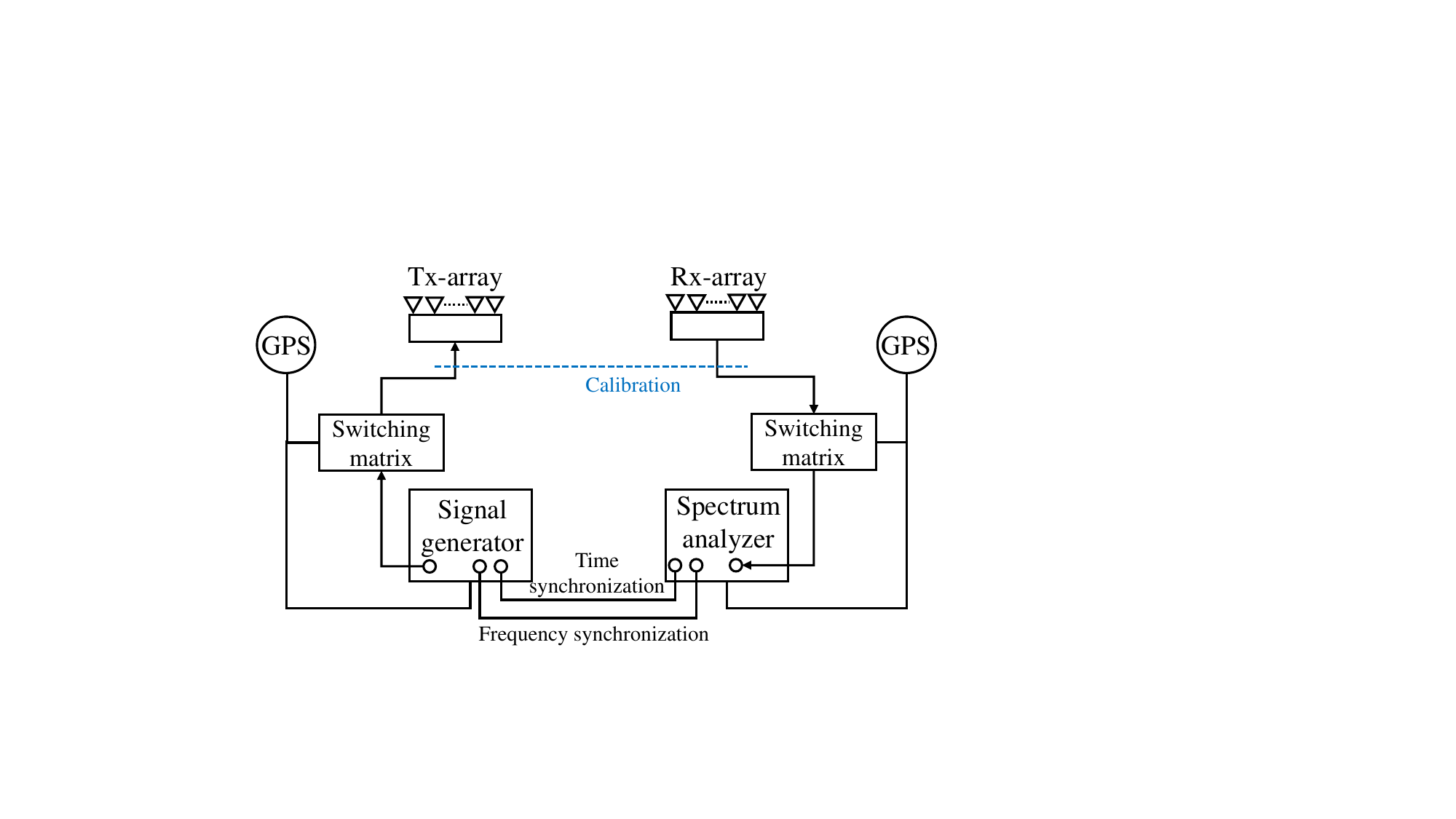}
	\caption{Schematic diagram of massive MIMO time-domain channel sounder.}
	\label{fig:Figure2}
\end{figure}

\par Table \uppercase\expandafter{\romannumeral1} shows the channel sounder and parameters. The transmitter end is composed of high frequency signal generator, control computer, switch matrix, GPS rubidium clock module, and array antenna. The power amplifier built into the switching matrix is used to increase the transmitting power and extend the coverage range of measurement signals. The receiving end consists of array antenna, spectrum analyzer, control computer, switch matrix, GPS rubidium clock module, and measurement software. The received signal is written as

\begin{equation}
\begin{split}
\mathbf{Y}(t)& = \sum_{l=1}^{L}exp(j2\pi \gamma_{l} t)\mathbf{C}_2(\mathbf{\Omega}_{2,l}\mathbf{A}_l\mathbf{C}_1^T(\mathbf{\Omega}_{1,l})u(t-\tau_l) 
\\& \quad + \sqrt{\dfrac{N_0}{2}}\mathbf{N}(t),
\end{split} 
\end{equation} 
\\ where $\gamma_{l}$ is Doppler frequency, $\mathbf{A}_l$ is amplitude, $\tau_l$ is delay of the $l$th multipath component (MPC). $(\cdot)^T$ is the transpose operator, $\mathbf{N}(t)$ is the standard complex white Gaussian noise with power spectral density $N_0$. $\mathbf{C}_1(\mathbf{\Omega}_{1,l})$ and $\mathbf{C}_2(\mathbf{\Omega}_{2,l})$ are the steering vector of Tx array and Rx array, respectively.

\begin{table}[htbp]
	\centering
	\caption{Channel Sounder and Performance of Equipments.}
	\setlength{\tabcolsep}{0.1 mm}
	\label{my-label}
	\renewcommand{\arraystretch}{1.8}
	\setlength{\tabcolsep}{5 mm}
	\begin{tabular}{c|c}
		\hline \hline
		Equipment    & Parameters  \\ \hline
		\multirow{3}{*}{\makecell{Signal generator}}    &  Frequency range: 100 KHz-40 GHz \\ 
		\cline{2-2} & Maximum bandwidth: 2 GHz\\
		\cline{2-2} & Maximum output power: 25 dBm\\  \hline
		\multirow{3}{*}{\makecell{Spectrum analyzer}}    &  Frequency range: 2 Hz-43.5 GHz \\ 
		\cline{2-2} & Maximum bandwidth: 2 GHz\\
		\cline{2-2} & Maximum input power: 30 dBm\\  \hline
		\multirow{5}{*}{\makecell{Tx switch matrix}}    &  Frequency range: 3-16 GHz \\
		\cline{2-2} & Number of channels: 128\\ 
		\cline{2-2} & Overall gain: $\geq$ 27 dB\\
		\cline{2-2} & Switching time: $\leq$ 100 ns\\ 
		\cline{2-2} & Isolation: $\geq$ 40 dB\\ \hline
		\multirow{5}{*}{\makecell{Rx switch matrix}}    &  Frequency range: 3-16 GHz \\
		\cline{2-2} & Number of channels: 64\\ 
		\cline{2-2} & Overall gain: $\geq$ 33 dB\\
		\cline{2-2} & Switching time: $\leq$ 100 ns\\ 
		\cline{2-2} & Isolation: $\geq$ 40 dB\\ \hline
		\multirow{2}{*}{\makecell{Synchronous mode}}    &  GPS rubidium clock  \\
		\cline{2-2} & Precision: nanosecond\\ \hline
		\multirow{4}{*}{\makecell{Array antenna}}    &  Array angular resolution: $\leq$ 2$^\circ$  \\
		\cline{2-2} & Polarization: Custom, e.g., $\pm45^\circ$\\ 
		\cline{2-2} & Type: Custom, e.g., UPA / ODA\\ 
		\cline{2-2} & Frequency: Custom, e.g., 6 GHz\\ \hline
		\hline
	\end{tabular}
\end{table}

\par In Fig. 2, the transceiver antenna matrix module uses trigger signal and reference clock provided by GPS to realize high precision synchronization start and synchronization unit switching. The signal transceiver uses the reference clock provided by GPS to ensure the consistency of baseband signal at the transmitting end, signal carrier frequency and I/Q demodulation frequency at the receiving end, and time sequence of time domain sampling. In addition, the sampling trigger signal is used to realize the synchronization between the initial sampling time of the signal receiving device and the switching start time of the transceiver antenna matrix module. As a whole, the high accuracy synchronization of signal transmission, antenna switching and signal acquisition are realized.

	
	

\subsection{ Measurement Data Post-Processing}

\subsubsection{Channel Impulse Response}

\par To remove the impact of systems on channel characteristics and get real channel impulse response (CIR), a back-to-back calibration should be made before conducting measuring. The received signal can be expressed as

 \begin{equation}
\begin{split}
y(\tau)& = x(\tau) * h_{Tx}(\tau) * a_{Tx}(\tau) * h(\tau) * a_{Rx}(\tau) * h_{Rx}(\tau),
\end{split} 
\end{equation}
\\ where
``$*$'' denotes convolution in the time domain, \emph{$x(\tau)$} is the transmitted signal. $h(\tau)$ is the CIR. $h_{Tx}(\tau)$ and $h_{Rx}(\tau)$ are the system response functions of the Tx and Rx equipment, e.g., signal generator, spectrum analyzer, amplifier, and cables. $a_{Tx}(\tau)$ and $a_{Rx}(\tau)$ are the antenna response functions of Tx and Rx antenna. The calibration signal can be expressed as

\begin{equation}
\begin{split}
y_{cal}(\tau)& = x(\tau) * h_{Tx}(\tau) * h_{Rx}(\tau).
\end{split} 
\end{equation}

\par  By taking the Fourier transform of the above equations and assuming that the system is linear, we can get the ratio of the received signal data to the calibration signal data as

\begin{equation}
\begin{split}
\dfrac{Y(f)}{Y_{cal}(f)}& = \dfrac{X(f)H_{Tx}(f)A_{Tx}(f)H(f)A_{Rx}(f)H_{Rx}(f)}{X(f)H_{Tx}(f)H_{Rx}(f)} .
\end{split} 
\end{equation}

 The antenna response can be obtained by measuring in an anechoic chamber. The CIR is then obtained from inverse Fast Fourier transform as 

\begin{equation}
\begin{split}
h(\tau)& = IFFT(H(f)) = IFFT\left(\dfrac{Y(f)}{Y_{cal}(f)A_{Tx}(f)A_{Rx}(f)}\right),
\end{split} 
\end{equation} 
\\ where
$IFFT$ is the Inverse Fast Fourier Transform.

\subsubsection{SAGE Estimation Algorithm}

\par The high-resolution space-alternating generalized expectation-maximization (SAGE) algorithm is widely used for wireless channel parameter estimations. The algorithm can jointly estimate the complex amplitude, delay, azimuth angle, and elevation angle of the MPCs. The received signal $y(t)$ is assumed to consist of a finite number $L$ of propagation paths, i.e.,

\begin{equation}
\begin{split}
y(t)& = \sum_{l=1}^{L}s(t;\mathbf{\Theta}_l) + \sqrt{\dfrac{N_0}{2}}\mathbf{N}(t),
\end{split} 
\end{equation} 
\\ where
$s(t;\mathbf{\Theta}_l)$ is the is the $l$th MPC. The parameter sets to be
estimated are $\mathbf{\Theta}_l=[\mathbf{\Omega}_{1,l},\mathbf{\Omega}_{2,l},\tau_l,\gamma_l,\alpha_l]$, which are departure direction, arrival direction, propagation delay, doppler frequency and complex amplitude for the $l$th MPC, respectively. $\mathbf{N}(t)$ represents N-dimensional complex white Gaussian noise, which is $[N_1(t),...,N_N(t)]^T$, and $N_0$ is a positive number. The direction in which the wave arrives or leaves is described by the unit vector $\mathbf{\Omega}$, which can uniquely determine its spherical coordinates $\phi$ (azimuth angle) and $\beta$ (elevation angle), i.e

\begin{equation}
\begin{split}
\mathbf{\Omega}& = E(\phi, \beta) = [cos(\phi)sin(\beta), sin(\phi)sin(\beta), cos(\beta)]^T.
\end{split} 
\end{equation} 

\par The EM algorithm is introduced, because the log-likelihood function particular form of $\Lambda(\mathbf{\Theta}_l; \mathbf{x}_l)$,

\begin{equation}
\begin{split}
Q(\mathbf{\Theta}_l,\hat{\mathbf{\Theta}}^{'})& = \Lambda(\mathbf{\Theta}_l; \mathbf{x}_l).
\end{split} 
\end{equation} 
\par Then, for the maximization step, by setting an initial parameter vector $\hat{\mathbf{\Theta}}^{'}$ and iterating, we end up with an estimate $\hat{\mathbf{\Theta}}^{''}$. Such that it satisfies $\hat{\mathbf{\Theta}}^{''}=\mathrm{argmax}_{\mathbf{\Theta}_l}\left\{Q(\mathbf{\Theta}_l,\hat{\mathbf{\Theta}}^{'}) \right\}$. The parameter estimates can be updated iteratively by using the SAGE principle, which is a typical channel parameter estimation algorithm for MIMO measurements [33, 34].

\section{Channel Measurement and Characterization in Near-Field }

\par In the 6G era, with the increase of antenna scale, XL-MIMO near-field channel measurement has received extensive attention. There are two issues that need to be discussed, i.e., Can the near-field spatial non-stationary be observed along the array? Are there some model variation rules for the near-field SnS of channel characteristics?

\subsection{Near-field Property}

\par When the antenna arrays reach such a large dimension, SnS characteristics appear along the array. Different parts of the array may have different views of the propagation environment, or different channel paths. In the near-field region, the channel is likely to be modeled under the spherical wave assumption, where the phase and the difference in traveling distance at different antennas cannot be ignored. However, when the array is divided into the form of a subarray, the phase and the difference in traveling distance at different antennas can be ignored within the subarray. The channel satisfies the wide sense stationary uncorrelation (WSSU). In other words, this is a refined near- and far-field separation.

\par In this paper, the subarray of the antenna array is defined as the SI, where the large/small scale fading characteristics are stationary. At present, the division method of SI is usually uniform division based on the experience, but does not consider environmental factors [13, 35]. In the actual environment, the channel is dynamic due to the randomness of the scattering environment, that is, the channel matrix presents a nonlinear structure, and the SIs of antenna array are more likely to exhibit irregularity. Therefore, this paper proposes the potential adaptive division method based on channel characteristic.

\par To analyze the fluctuation of channel characteristic parameters $s \in [DS, AS, SF, K, \mathbf{\Omega}_{1},\mathbf{\Omega}_{2},\tau, P]$ on the array domain, some modeling methods can be used for the single parameter, such as the variance, range, coefficient of variation, etc. The slope $k$ of the model curve that the parameters $s$ on the array can be used for measurement.

\begin{equation}
\begin{split}
k & \triangleq \lim_{\Delta n \to 0} \dfrac{s(n+\Delta n)-s(n)}{\Delta n},
\end{split} 
\end{equation}
where DS, AS, SF, K and P are delay spread, angular spread, shadow fading, Ricean K-factor and power, respectively.

\begin{figure}[htbp]
	\setlength{\abovecaptionskip}{0.1 cm}
	\centering
	\includegraphics[width=0.45\textwidth]{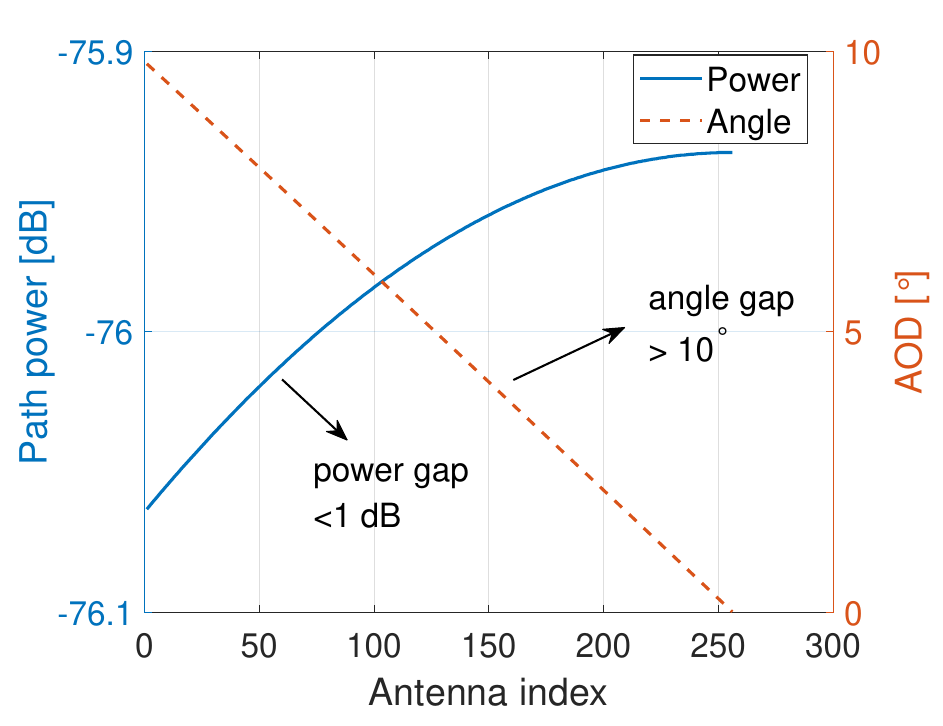}
	\caption{Near-field spherical-wave property}
	\label{fig:Figure3}
\end{figure}

\par Instead of phase variations as for Rayleigh distance, the uniform-power distance concerns the signal amplitude/power variations across array elements, where the power ratio between the weakest and strongest array element is no smaller than a certain threshold. We find that the angle on the array is constant in the far-field, so we discuss that the angle domain can better describe the difference between the near-field and the far-field. In our work, we can extract the angle of LOS paths from all receiving ends to all transmitting antenna array elements.

\par Fig. 3 shows that the maximum power difference is only 0.12 dB across array elements, and the change of power is not obvious in the array domain. The main reason is that the power/path loss is a large-scale parameter. However, the largest difference is obvious for angle with 10$^\circ$.

\subsection{Near-Field Channel Measurement}

\par The indoor is an important scenario for the application of XL-MIMO array. As shown in Fig. 4, a large classroom of the teaching building in Beijing University of Posts and Telecommunications (BUPT). The classroom is 17.1 m long, 11.5 m wide, and 3.0 m tall. The measurements are carried out in a indoor environment without the movement of people and other obstructions to keep the channel quasi-static. Based on the mechanical sliding platform, the small UPA array is translated to form a XL-MIMO array with a total of 1600 array elements. The ODA omnidirectional array is placed on the receiving end. There are some measurement routes in the classroom. Fig. 4 shows four typical measurement positions, distributed in different directions of the classroom for data comparative analysis of four positions. The equipment required for measurement and parameter settings are given in Table \uppercase\expandafter{\romannumeral2}.


\begin{figure}[htbp]
	\setlength{\abovecaptionskip}{0.1 cm}
	\centering
	\includegraphics[width=0.40\textwidth]{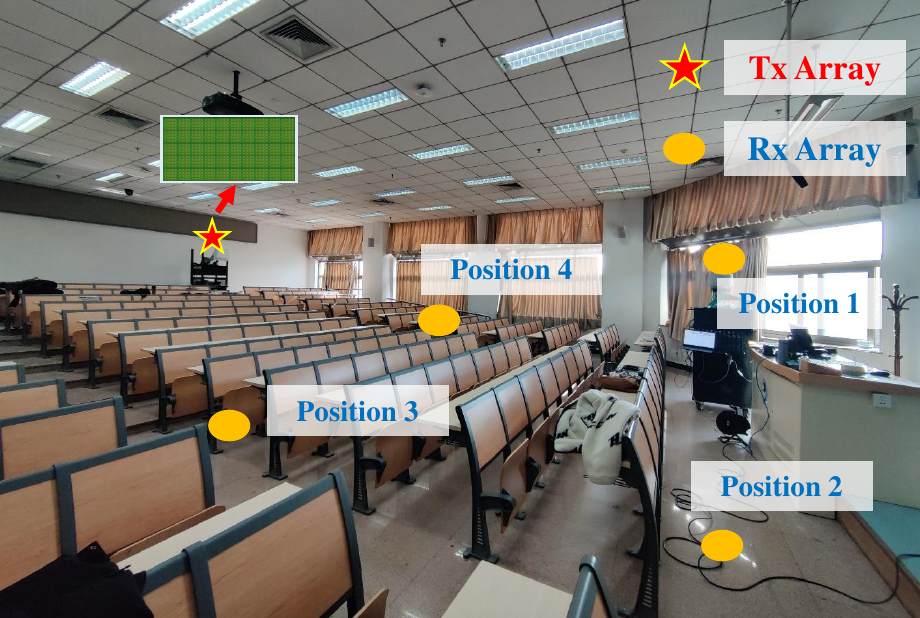}
	\caption{Near-field XL-MIMO channel measurement in the indoor scenario.}
	\label{fig:Figure5}
\end{figure}

\begin{table}[htbp]
	\centering
	\caption{Parameters of Massive MIMO Channel Sounding Platform}
	\setlength{\tabcolsep}{0.1 mm}
	\label{my-label}
	\renewcommand{\arraystretch}{1.5}
	\setlength{\tabcolsep}{12 mm}
	\begin{tabular}{c|c}
		\hline \hline
		Parameter    & Performance  \\ \hline
		Frequency [GHz]    & 6  \\ \hline
		Bandwidth [MHz]     & 200  \\  \hline
		Symbol rate [Mbaud]     & 100 \\  \hline
		Chip length     & 511 \\  \hline
		Tx antenna array  & UPA  \\  \hline
		Rx antenna array  & ODA \\  \hline
		Polarization & $\pm45^\circ$   \\   \hline \hline
	\end{tabular}
\end{table}

\subsection{Verification of Near-Field Channel Characteristics}	

\par This section will analyze the RMS delay spread, angular spread and power of LOS path on the array. The different fitting accuracy and brevity are compared, and polynomial is used for modeling, which can be expressed as

\begin{equation}
\begin{split}
y & = A \cdot n^2 + B \cdot n + C + N(0,\sigma^2),
\end{split} 
\end{equation}
where A, B and C are polynomial model coefficients, and $\sigma$ is the fit standard deviation.

\subsubsection{ Power and Angle of LOS Path}

\par This section explores the variation model on an array at different Rx locations, including power and angle of departure (AOD).

\begin{figure}[htbp]
	\xdef\xfigwd{\columnwidth}
	\setlength{\abovecaptionskip}{0.1 cm}
	\centering
	\begin{tabular}{c}
		\includegraphics[width=0.42\textwidth]{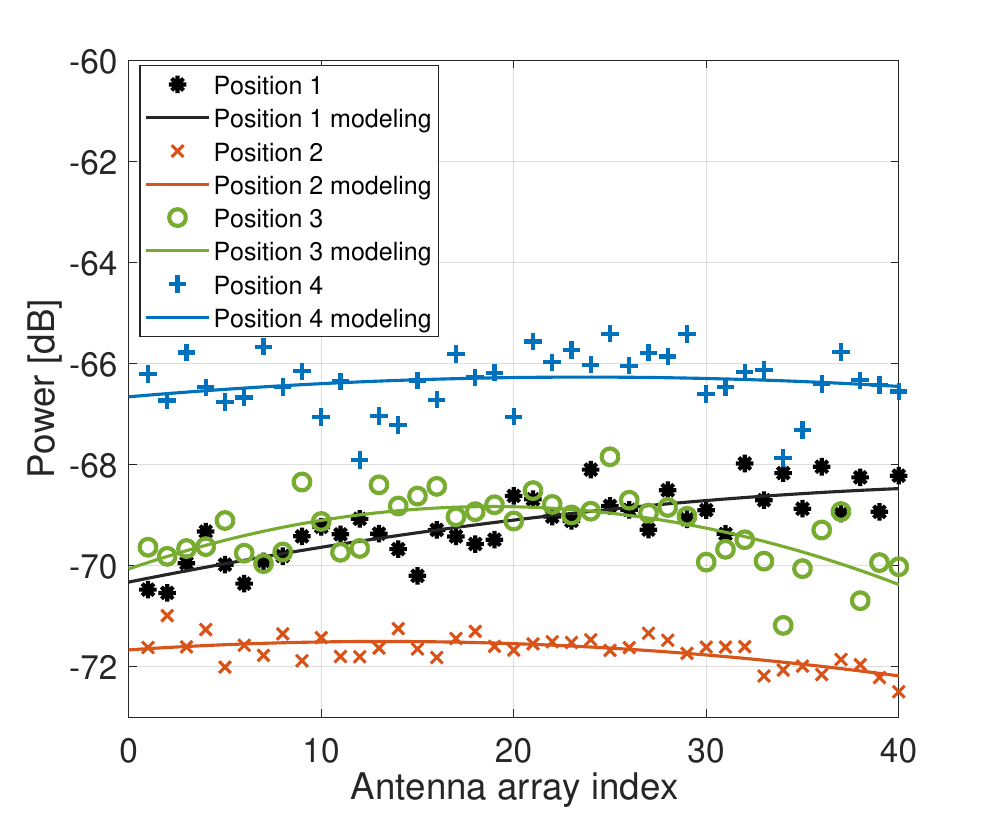}\\
		{\footnotesize\sf (a)} \\[3mm]
		\includegraphics[width=0.42\textwidth]{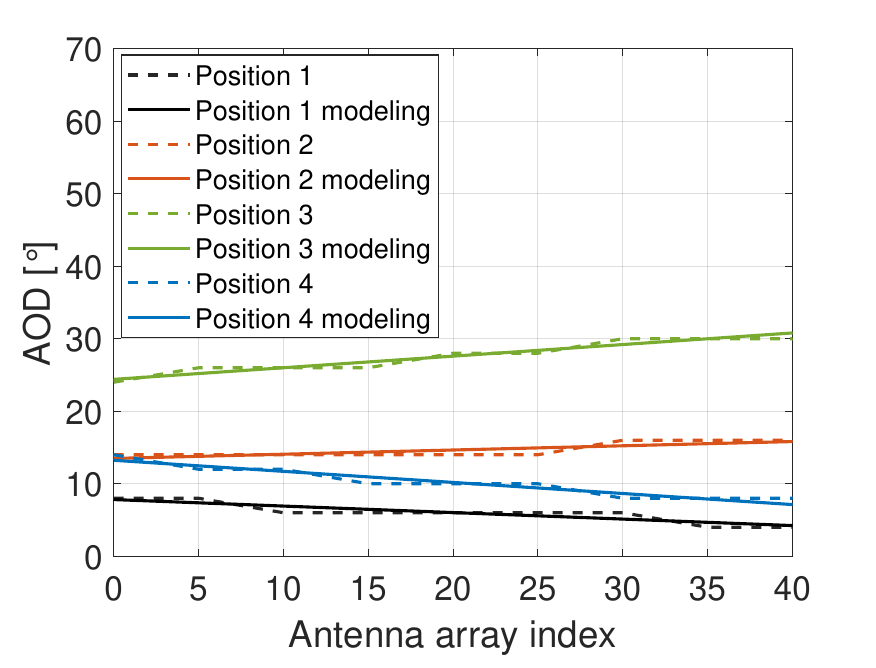}\\
		{\footnotesize\sf (b)} \\
	\end{tabular}
	\caption{Power and angle. (a) Power. (b) AOD for LOS path.}
	\label{fig:Figure6}
\end{figure}

\begin{table}[htbp]
	\centering
	\caption{Angle model parameters for LOS Path }
	\setlength{\tabcolsep}{0.1 mm}
	\label{my-label}
	\renewcommand{\arraystretch}{1.5}
	\setlength{\tabcolsep}{7 mm}
	\begin{tabular}{c|c|c|c}
		\hline \hline
		Position  & B& C & $\sigma$ \\ \hline
		Position 1    & -0.45   & 8.27 & 0.61             \\ \hline
		Position 2     & 0.29 & 13.2  & 0.57       \\  \hline
		Position 3     & 0.80     &23.6 &0.63          \\  \hline
		Position 4    & -0.76  & 14.0 & 0.66               \\  \hline
		\hline
	\end{tabular}
\end{table}

\par As shown in Fig. 5(a), along the horizontal domain of the array, when linear array elements change, the power changes in a fluctuation range of about 1-2 dB. Following the quadratic polynomial model, the LOS power on the array element at one end of the transmitting antenna array closer to the receiving antenna is almost stable. However, the power of other transmitting array which is far away from the receiving antenna will change obviously with the change of array position. As shown in Fig. 5(b) and Table \uppercase\expandafter{\romannumeral3}, the LOS angle presents a linear change with the change of the array elements, which is consistent with the observation in the simulation, indicating the occurrence of SnS phenomenon. By comparing different positions, it can be found that the deviation angle between position 3 and the antenna array is larger, and the non-stationary fluctuation degree is larger (the slope is the largest), indicating that the spatial non-planarity has a great relationship with the offset angle between the receiving point and the base station array. At the same time, the absolute slope of position 4 is also large, because position 4 is closer to the launching end. The results show that when the SI is divided, the SI judged by amplitude is larger than that by angle or phase.

\par In addition, it is found through data comparison that when the influence of NLOS multipaths is considered, the variation trend of the total received power of the array elements will change greatly on the array. The results show that the abundance of multipaths in near-field affects the tendency of SnS properties compared to considering only LOS path.

\subsubsection{Delay and Angular Domain}

\par In Fig. 6(a), the delay spread on the array will have relatively large fluctuations, which is due to the abundant indoor multipaths. Under the action of rich multipath, the difference between different Rx positions is quite large. The results show that the receiver and receiver positions (distance and angle) have a great influence on the near-field channel characteristics. In Table \uppercase\expandafter{\romannumeral4}, the polynomial model can be used to generate the delay spread on different array elements, and the fitting accuracy of the quadratic model is higher, because the power changes in the array domain is quadratic. In addition, the delay spread can also be generated by using a first-order polynomial fitting model.

\begin{figure}[htbp]
	\xdef\xfigwd{\columnwidth}
	\setlength{\abovecaptionskip}{0.1 cm}
	\centering
	\begin{tabular}{c}
		\includegraphics[width=0.42\textwidth]{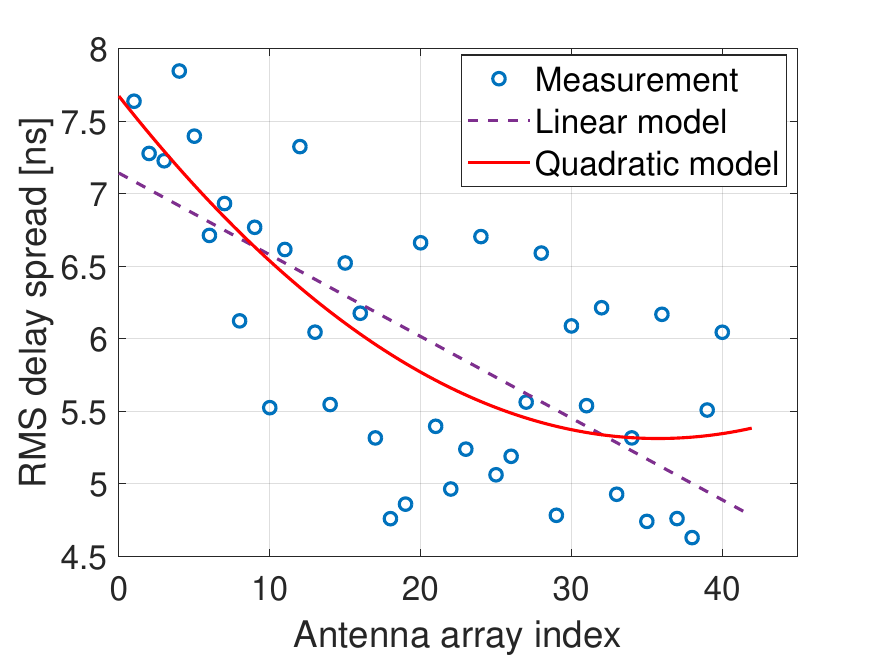}\\
		{\footnotesize\sf (a)} \\[3mm]
		\includegraphics[width=0.42\textwidth]{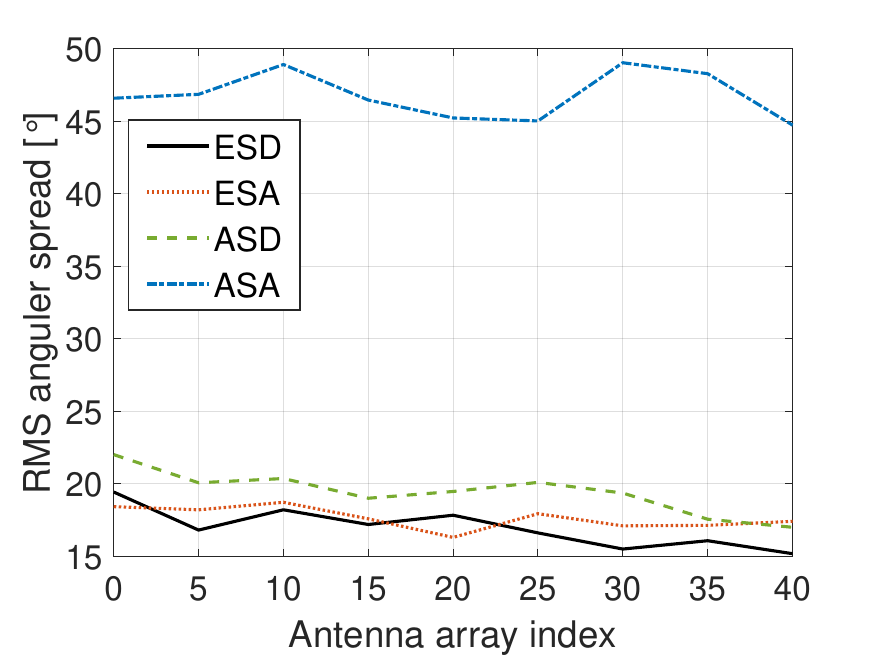}\\
		{\footnotesize\sf (b)} \\
	\end{tabular}
	\caption{ Near-field region. (a) RMS delay spread . (b) RMS angular spread.}
	\label{fig:Figure7}
\end{figure}

\par As shown in Fig. 6(b) and Table \uppercase\expandafter{\romannumeral4}, ASA, ESA, ASD, and ESD are angular spreads of the azimuth angle of arrival, the elevation angle of arrival, the azimuth angle of departure, and the elevation angle of departure, respectively. The linear model can be used to fit the angular spread in the array domain. Most azimuth angular spreads are larger than elevation angular spreads because the classroom scenario is more open horizontally. The ASA on the array is relatively larger. The main reason is that the receiving end is omnidirectional array, which can receive more effective multipaths in the spatial domain.

\begin{table}[htbp]
	\centering
	\caption{Delay spread and angular spread model parameters.}
	\setlength{\tabcolsep}{0.1 mm}
	\label{my-label}
	\renewcommand{\arraystretch}{1.5}
	\setlength{\tabcolsep}{2.1 mm}
	\begin{tabular}{c|c|c|c|c|c}
		\hline \hline
		Parameter    & Model & A & B& C & $\sigma$ \\ \hline
		\multirow{2}{*}{\makecell{Delay spread [ns]}}    &  Linear &  0&  -0.06&  7.14 & 0.74 \\ 
		\cline{2-6} &  Quadratic &  0.0018 &  -0.1321&  7.673 & 0.71  \\  \hline
		\multirow{2}{*}{\makecell{ESD [$^\circ$]}}    &  Linear &  0&  -0.39&  18.97 & 0.73 \\ 
		\cline{2-6} &  Quadratic &  0.0117 &  -0.5183&   19.22 & 0.78  \\  \hline
		
		\multirow{2}{*}{\makecell{ESA [$^\circ$]}}    &  Linear &  0&  -0.18&  18.53 & 0.60 \\ 
		\cline{2-6} &  Quadratic &  0.0168 &  -0.366 &  18.90 & 0.63  \\  \hline
		
		\multirow{2}{*}{\makecell{ASD [$^\circ$]}}    &  Linear &  0&  -0.51&  21.92 & 0.76 \\ 
		\cline{2-6} &  Quadratic &  -0.0228 &  -0.254&  21.42 & 0.79  \\  \hline
		
		\multirow{2}{*}{\makecell{ASA [$^\circ$]}}    &  Linear &  0&  -0.24&  47.75  & 1.84 \\ 
		\cline{2-6} &  Quadratic &  -0.0811 &  0.6561&  45.97 & 1.85  \\  \hline
		\hline
	\end{tabular}
\end{table}

\subsection{Spatial Non-Stationary}

\par This section explores and discusses SnS characteristics for the Tx array antenna at measurement positions. The RMS delay spreads at 6 GHz for 1600-element planar arrays are given in Fig. 7, with horizontal and vertical indices of the antenna positions in the virtual array.

\begin{figure}[htbp]
	\xdef\xfigwd{\columnwidth}
	\setlength{\abovecaptionskip}{0.1 cm}
	\centering
	\begin{tabular}{cc}
		\includegraphics[width=4.2cm,height=3.4cm]{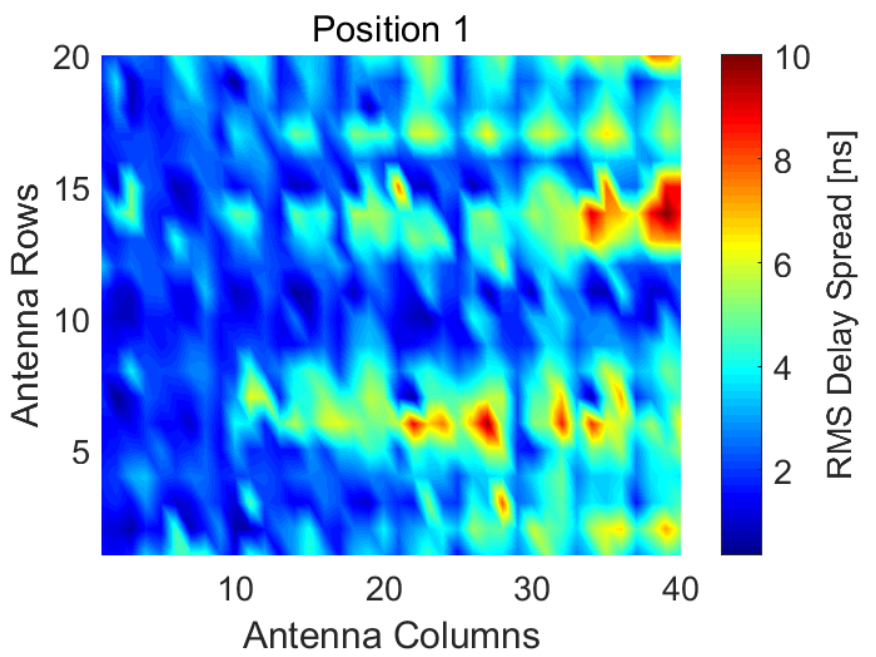} \hspace{-4mm}
		& \includegraphics[width=4.2cm,height=3.4cm]{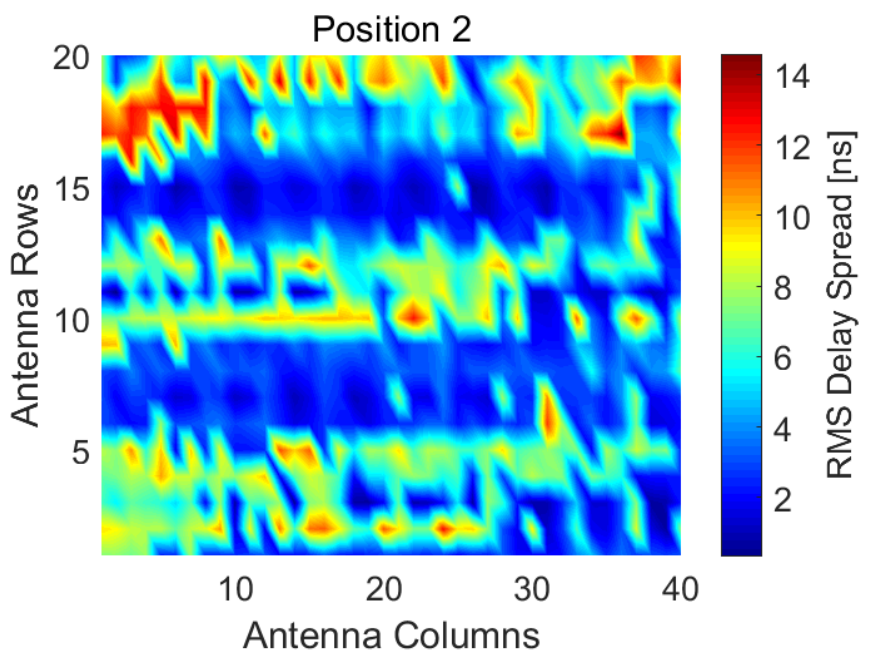}\\
		{\footnotesize\sf (a)} &	{\footnotesize\sf (b)} \\
		
	\end{tabular}
	\begin{tabular}{cc}
		\includegraphics[width=4.2cm,height=3.4cm]{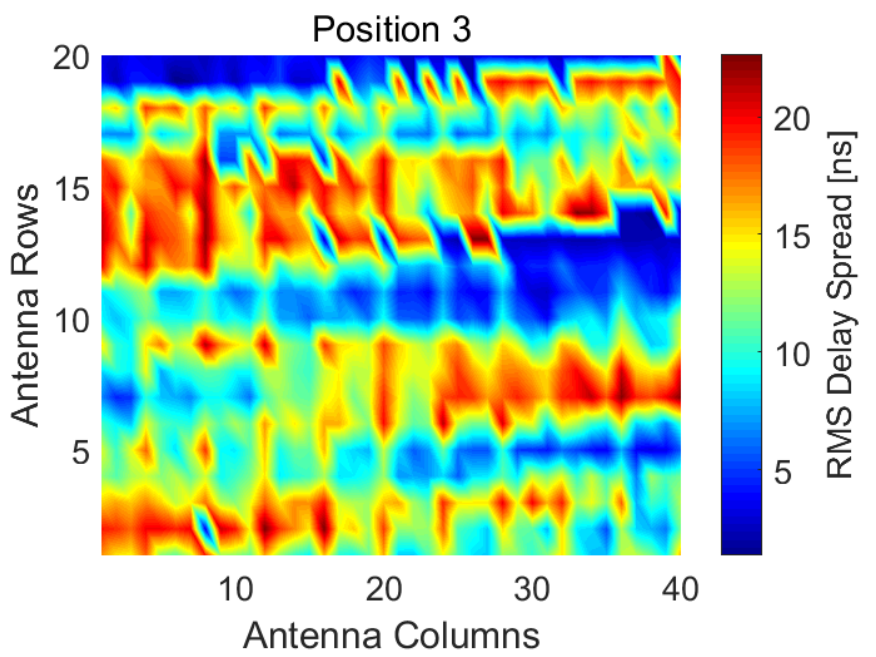} \hspace{-4mm}
		& \includegraphics[width=4.2cm,height=3.4cm]{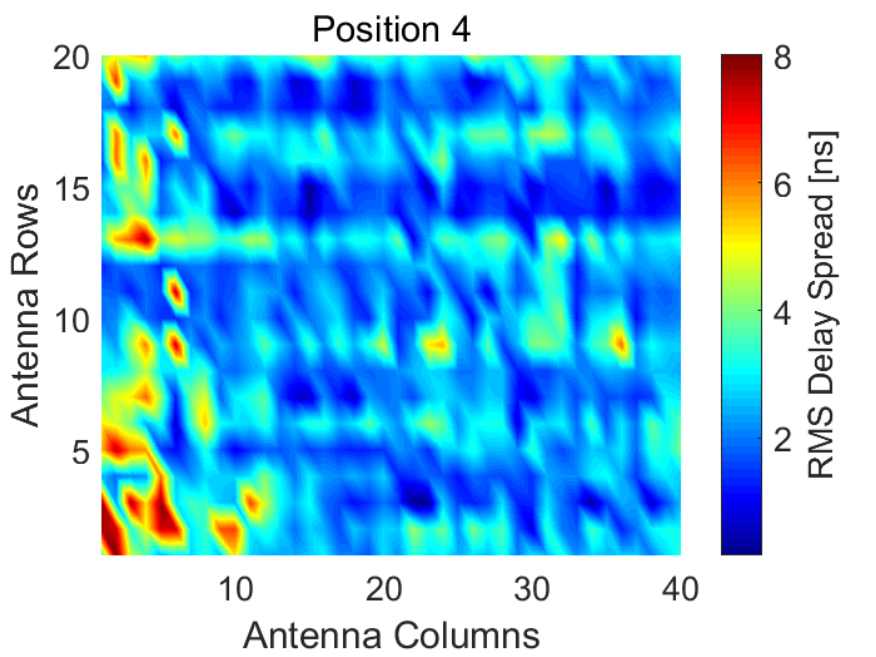}\\
		{\footnotesize\sf (c)} &	{\footnotesize\sf (d)} \\
	\end{tabular}
	\caption{ RMS delay spread for 1600-element planar array, with
		horizontal and vertical indices of the antenna positions on the virtual array. (a) Position 1. (b) Position 2. (c) Position 3. (d) Position 4.}
	\label{fig:Figure8}
\end{figure}

\par It can be observed from Fig. 7 that the SnS characteristics of RMS delay spreads at different receiving positions are significantly different. Therefore, it can be judged that the SnS of RMS delay spreads of indoor scenarios on the array needs to collect data of multiple receiving positions for statistical analysis. In Fig. 7, the RMS delay spread generally fluctuates in a range from 6 ns to 20 ns. Another fact is that the RMS delay spread varies on the array. The reason is that the power of the multipath components changes on the array, resulting in a different delay dispersion in each sub-channel. As shown in Fig. 7, compared with the four positions, the maximum range of delay spreads change on the array is about 15 ns. These observations of significant changes in RMS delay spread may be due to several reasons: i) the scatterer generated path undergoes the birth and death process from the subarray to the next subarray. ii) Shadow fading occurs throughout the array such that the path experiences significant attenuation when viewed at different positions.

\section{Channel Measurement and Characterization From Near-Field to Far-Field }

\par With the increase of antenna scale and frequency, the near-field range gradually expands, and the wireless communication environment may have both near-field and far-field regions. There is the issues that need to be discussed, i.e., Are there the SnS model of channel characteristics from near-field to far-field?

\subsection{Channel Measurement}

\par In the open space in front of the teaching building of the Beijing University of Posts and Telecommunications in Fig. 8. The UPA array with 32 elements at the origin is translated to form a 1 m long ultra-large virtual array, with a total of 320 elements. The ODA three-dimensional omnidirectional array with 56 array elements has 16 different receiving positions. The measurement work is divided into two cases in Fig. 9. Case A is a near-field channel measurement consisting of two routes. Route 1 is a straight line with a total of 16 receiving points spaced 1 m apart. Route 2 is a circular arc with a total of 7 receiving points and an angle of 20$^\circ$ between adjacent points. Case B is the channel measurement activity from near-field to far-field. The measurement settings are the same as that in Table \uppercase\expandafter{\romannumeral2}.


\begin{figure}[htbp]
	\setlength{\abovecaptionskip}{0.1 cm}
	\centering
	\includegraphics[width=0.40\textwidth]{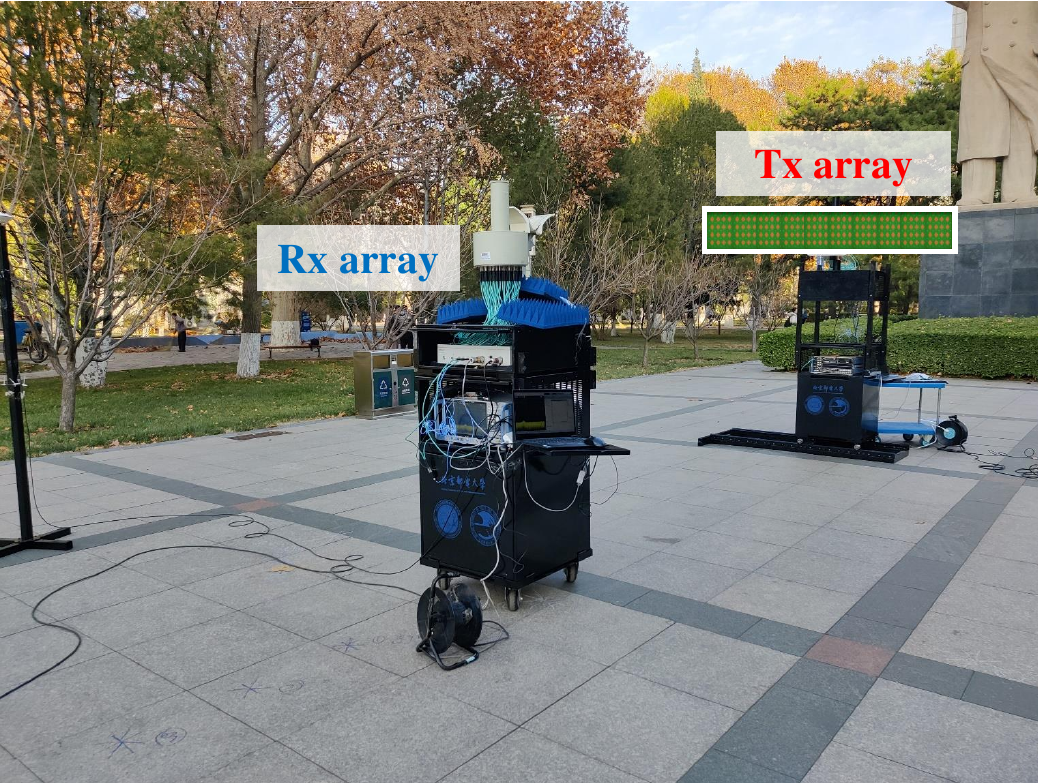}
	\caption{Near-field channel measurement in the outdoor scenario.}
	\label{fig:Figure8}
\end{figure}

\begin{figure}[htbp]
	\setlength{\abovecaptionskip}{0.1 cm}
	\centering
	\includegraphics[width=0.45\textwidth]{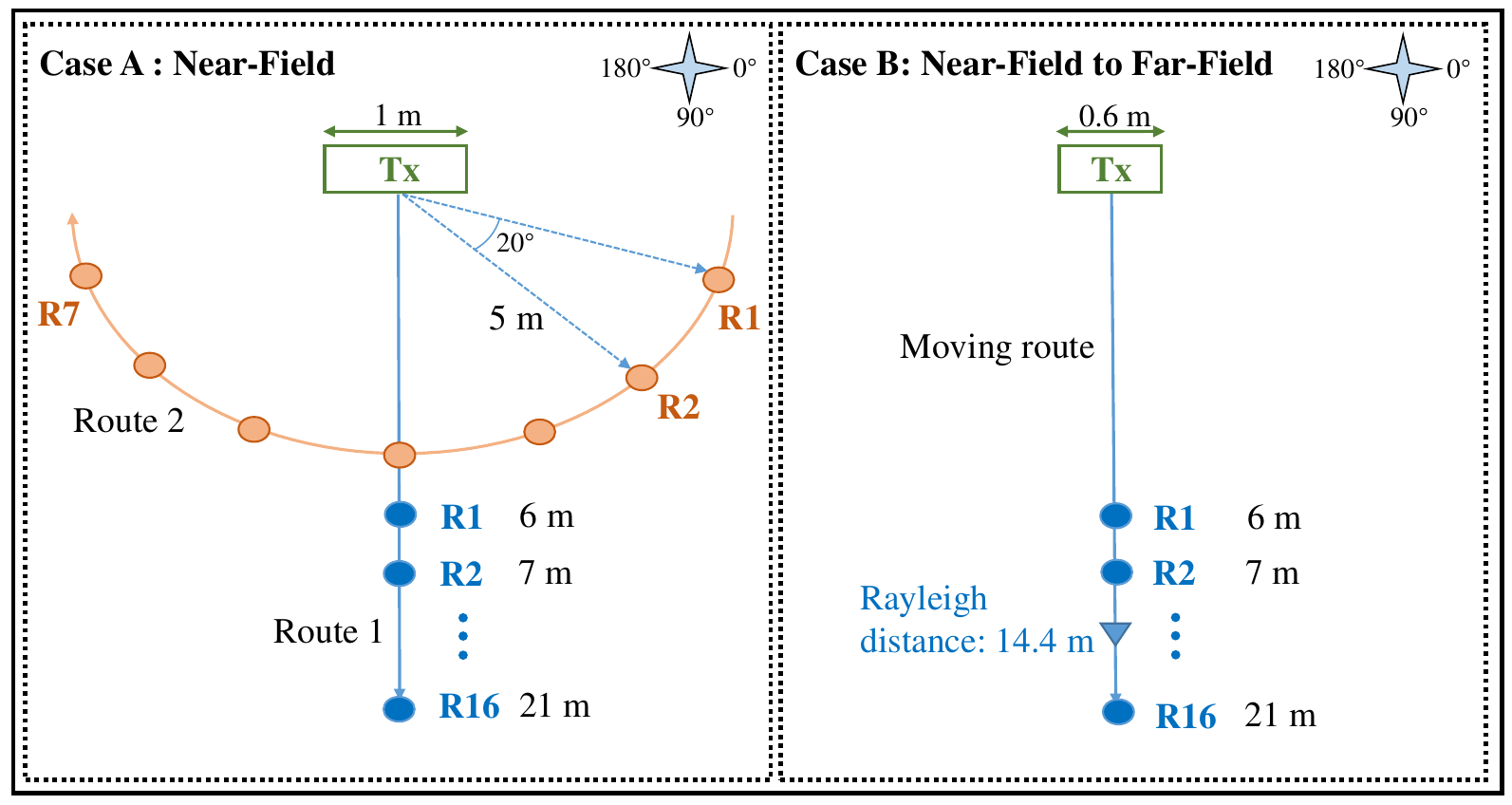}
	\caption{Channel measurement scheme and layout.}
	\label{fig:Figure9}
\end{figure}

\subsection{Channel Characteristics and Modeling}

\par This section explores the variation model on the array, including path loss, delay and angle.

\subsubsection{Path Loss}

\par The path loss is a fundamental channel model for estimating the link budget and coverage in cellular network. This section will focus on characterizing the spatial non-stationarity and path loss model in near-field and far-field.

\begin{figure}[htbp]
	\xdef\xfigwd{\columnwidth}
	\setlength{\abovecaptionskip}{0.1 cm}
	\centering
	\begin{tabular}{c}
		\includegraphics[width=0.45\textwidth]{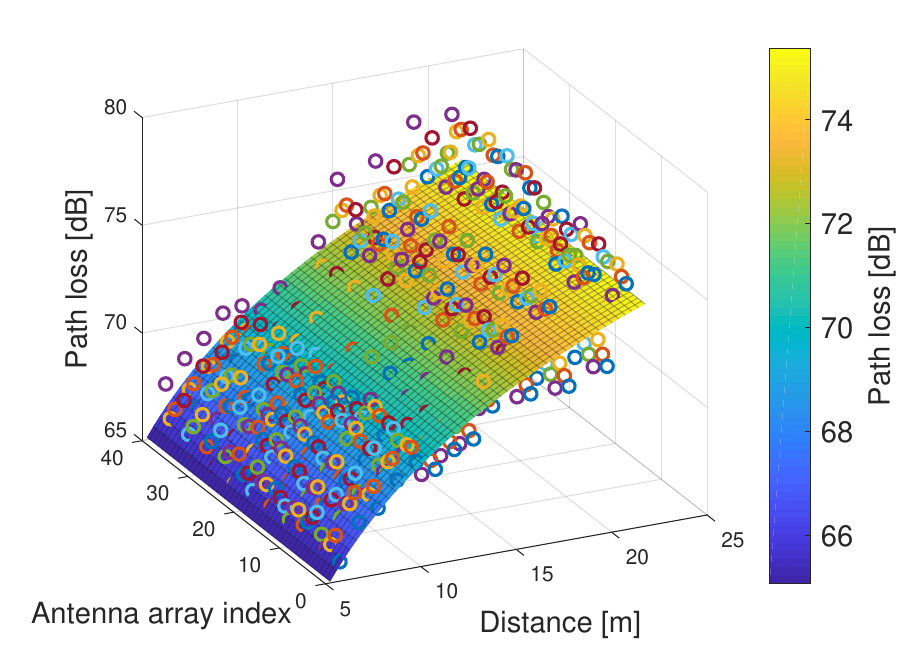}\\
		{\footnotesize\sf (a)} \\[3mm]
		\includegraphics[width=0.45\textwidth]{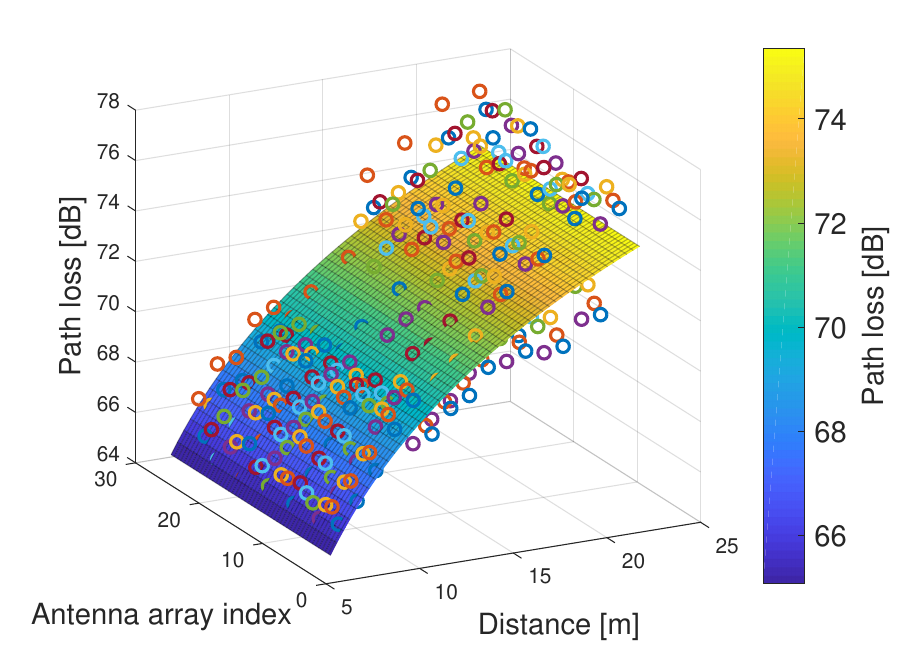}\\
		{\footnotesize\sf (b)} \\[3mm]
		\includegraphics[width=0.45\textwidth]{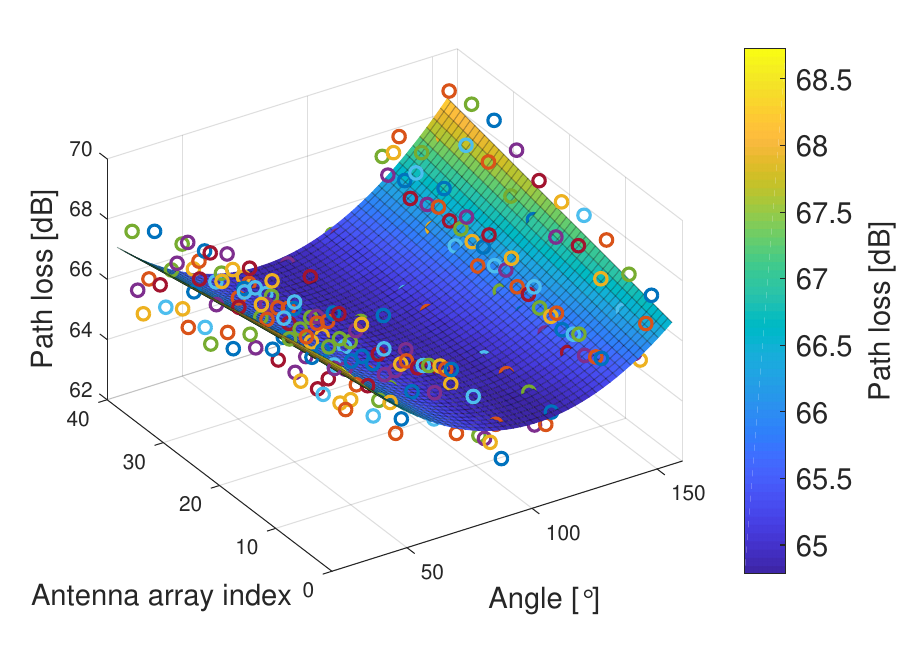}\\
		{\footnotesize\sf (c)} \\
	\end{tabular}
	\caption{Path loss. (a) Near-field (Route 1 in Case A). (b) From near-field to far-field (Case B). (c) Near-field (Route 2 in Case A).}
	\label{fig:Figure10}
\end{figure}

\par The different fitting accuracy and brevity of path loss model are compared, which can be expressed as

\begin{equation}
\begin{split}
PL & = A \cdot log_{10}(d) + B \cdot n + C + N(0,\sigma^2),
\end{split} 
\end{equation}
where $d$ and $n$ are the distance and antenna index, respectively. A, B and C are polynomial model coefficients, $\sigma$ is the fit standard deviation.

\begin{table}[htbp]
	\centering
	\caption{Path loss model parameters for near-field and far-field}
	\setlength{\tabcolsep}{0.1 mm}
	\label{my-label}
	\renewcommand{\arraystretch}{1.5}
	\setlength{\tabcolsep}{5 mm}
	\begin{tabular}{c|c|c|c|c}
		\hline \hline
		Scenario  & A& B& C & $\sigma$ \\ \hline
		NF & 16.60  & 0.0025   & 53.11 & 1.42             \\ \hline
		NF-FF  & 16.61   & 0.0009 & 53.14  & 1.46       \\  \hline
		\hline
	\end{tabular}
\end{table}

\begin{equation}
\begin{split}
PL & = 71.47 - 0.13\theta - 0.059n + 0.00065\theta^2 + 0.00066\theta n 
\\&\quad + N(0, 0.76^2),
\end{split} 
\end{equation}
where $\theta$ and $n$ are the angle and antenna index, respectively.

\par As shown in Fig. 10, as the array size increases, a near-field effect occurs. As shown in Fig. 10(a) and Fig. 10(b), whether the receiving end is in the near field range or moving from near-field to far-field, the path loss follows a logarithmic distribution with the change of distance from the center of Rx-Tx, but a linear distribution over the array domain. However, it can be seen from Table \uppercase\expandafter{\romannumeral5} that the spatial non-stationary phenomenon on the array becomes weaker during the change from near-field to far-field, and the absolute value of the slope on the array domain becomes smaller. In Fig. 10(c), when the Rx-Tx center distance is unchanged, the spatial non-stationary phenomenon of path loss observed by the receiving end at different directions is different, and the non-stationary phenomenon at both ends is more obvious, mainly because the difference between the distance between the receiving end at these positions and different elements is larger.

\par In addition, from Fig. 10(c) and model (11), it can be found that the distribution on the array field follows a linear distribution, but the relationship with the Angle is more complex and follows a quadratic function. The results suggest that the near-field range may not be a semicircular circle calculated by the classical Rayleigh distance directly, but may be more consistent with the property of an ellipse.

\subsubsection{Delay Domain}

\par This section explores and discusses SnS characteristics for the Tx array antenna at measurement positions. The RMS delay spreads for 320-element planar arrays are given in Fig. 11.

\begin{figure}[htbp]
	\setlength{\abovecaptionskip}{0.1 cm}
	\centering
	\includegraphics[width=0.42\textwidth]{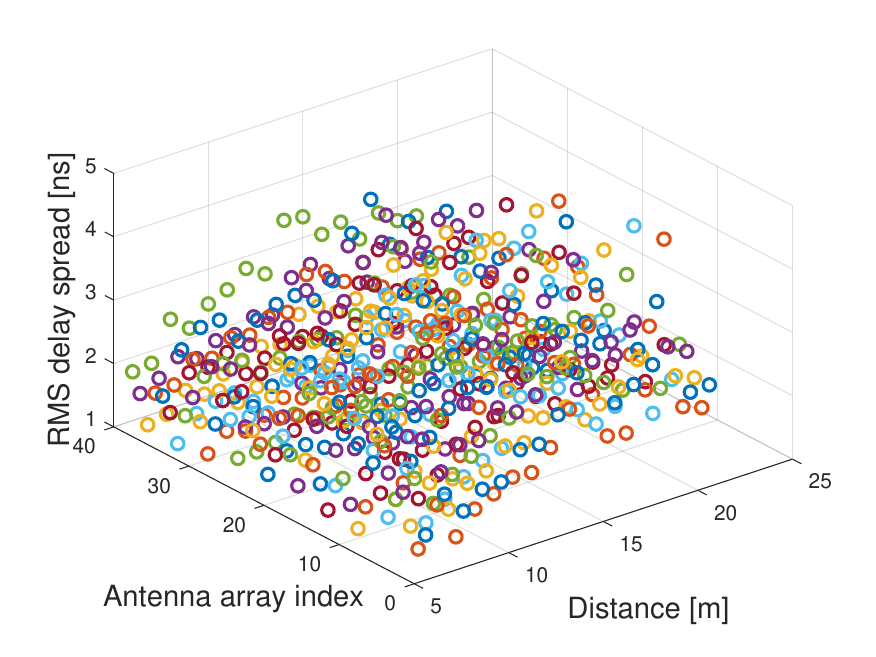}
	\caption{The delay spread in near-field (Route 1 in Case A).}
	\label{fig:Figure11}
\end{figure}

\par As shown in Fig. 11, in the near-field range, the delay spread does not show a significant change trend with the change of the distance of the Rx-Tx center. The results show that the relationship between delay expansion and distance is not clear, which is similar to the research results in the far field, that is, no model relationship between delay expansion and distance is established. However, through data fitting analysis on the array domain, it can be found that there is a certain relationship, but the difference between the linear model and the quadratic model is not large, probably because the size of the array is not large enough.

\begin{figure}[htbp]
	\xdef\xfigwd{\columnwidth}
	\setlength{\abovecaptionskip}{0.1 cm}
	\centering
	\begin{tabular}{cc}
		\includegraphics[width=4.2cm,height=3.4cm]{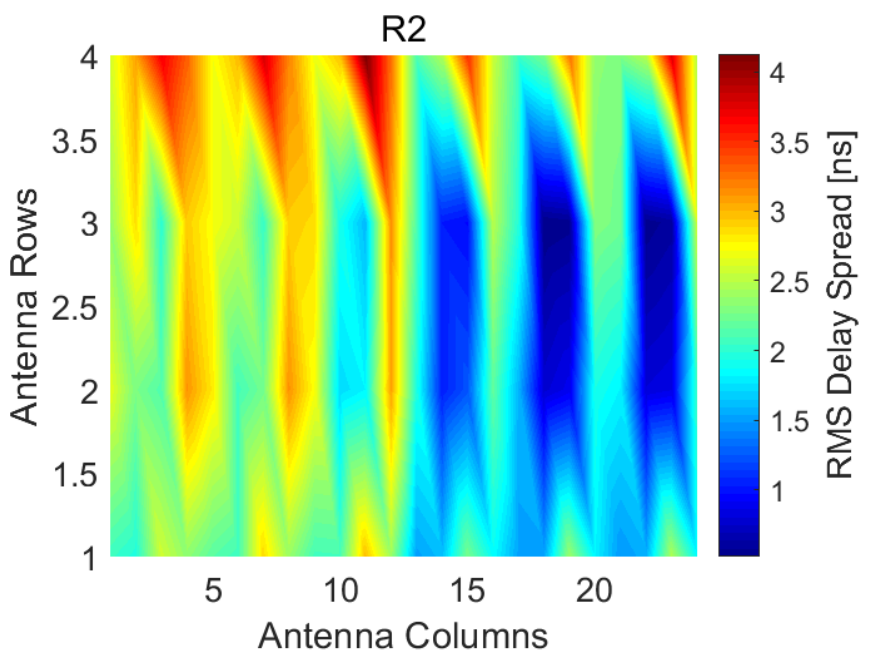} \hspace{-4mm}
		& \includegraphics[width=4.2cm,height=3.4cm]{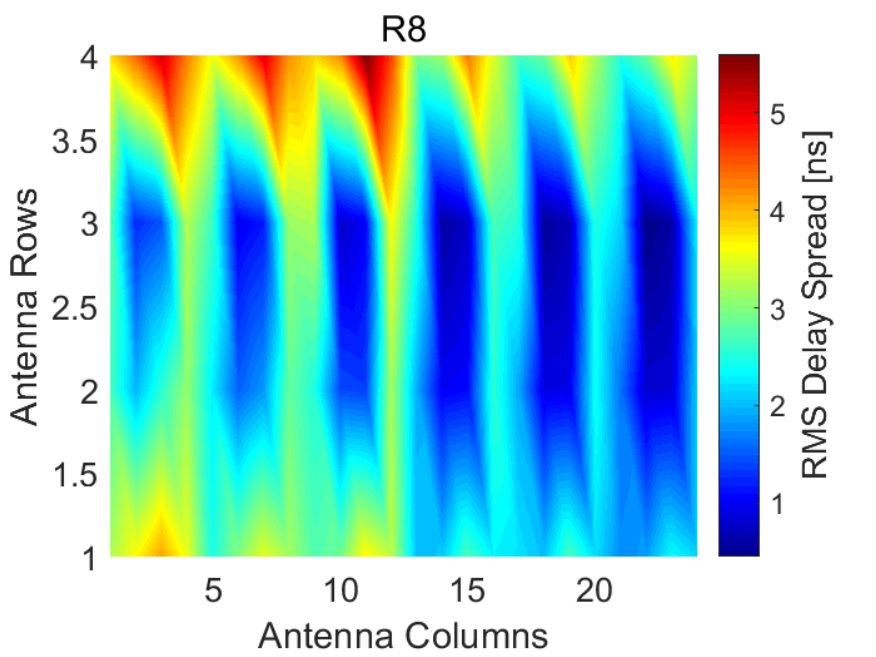}\\
		{\footnotesize\sf (a)} &	{\footnotesize\sf (b)} \\
		
	\end{tabular}
	\begin{tabular}{cc}
		\includegraphics[width=4.2cm,height=3.4cm]{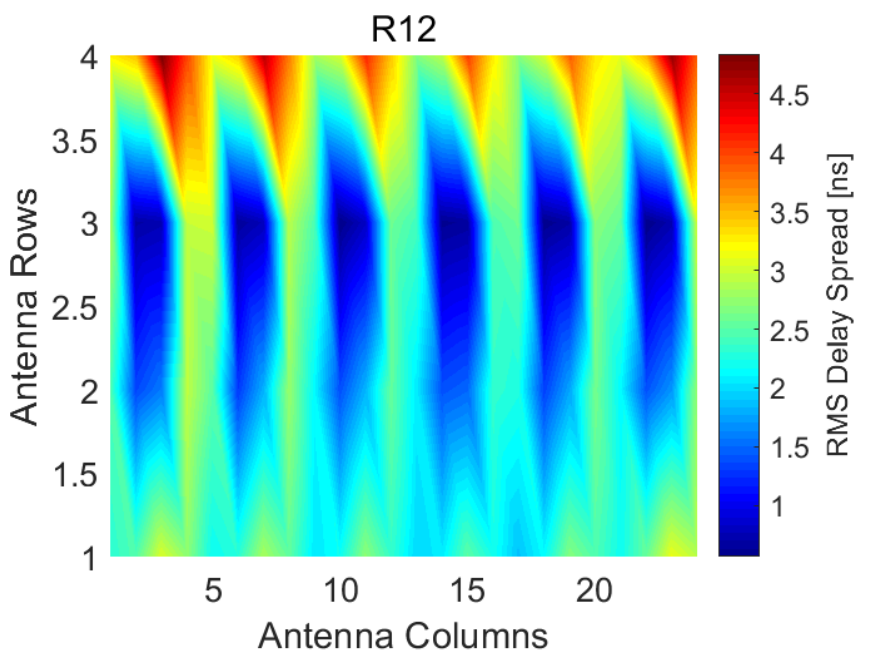} \hspace{-4mm}
		& \includegraphics[width=4.2cm,height=3.4cm]{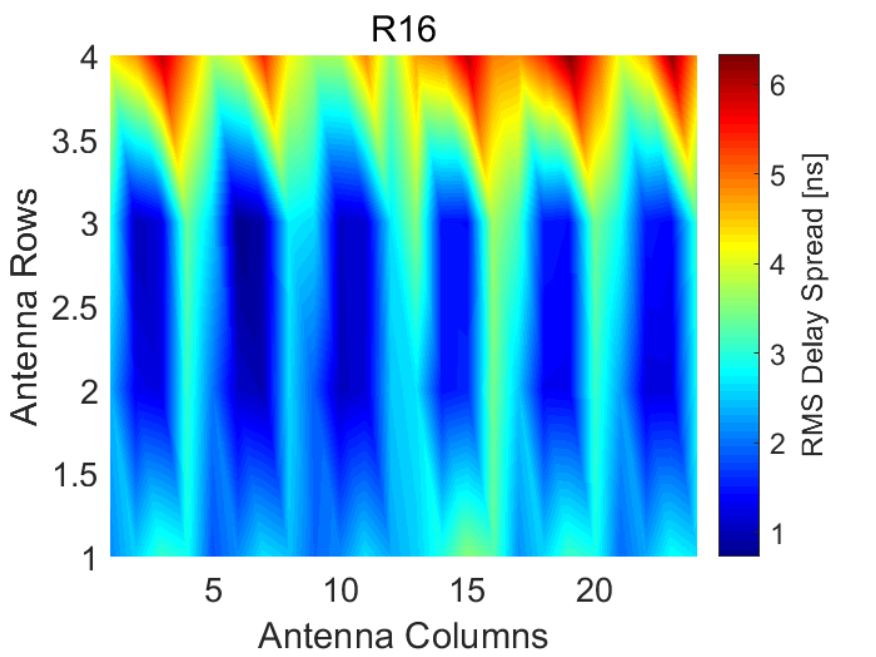}\\
		{\footnotesize\sf (c)} &	{\footnotesize\sf (d)} \\
	\end{tabular}
	\caption{ RMS delay spread from near-field to far-field in Case B. (a) R2 in near-field. (b) R8 in near-field. (c) R12 in far-field. (d) R16 in far-field.}
	\label{fig:Figure12}
\end{figure}

\par As shown in Fig. 12, from near-field to far-field, the space of delay spread on the array changes significantly. In Fig. 12(a), the position of Rx is closer to the array antenna, and the horizontal dimension of the array can be seen to be obviously non-stationary, with the maximum difference exceeding 3 ns. As Rx moves away, the non-stationary phenomenon in Fig. 12(b) weakens, with a maximum difference of about 2 ns. By comparison with Fig. 12(c) and (d), Rx has entered the far-field range, and the non-stationary phenomenon on the array domain has become very obvious.

\subsubsection{Spatial Domain}

\par As shown in Fig. 13, Fig. 14, and Table \uppercase\expandafter{\romannumeral6}, ASA, ESA, ASD, and ESD are angular spreads of the azimuth angle of arrival, the elevation angle of arrival, the azimuth angle of departure, and the elevation angle of departure, respectively.

\begin{figure}[htbp]
	\xdef\xfigwd{\columnwidth}
	\setlength{\abovecaptionskip}{0.1 cm}
	\centering
	\begin{tabular}{cc}
		\includegraphics[width=4.2cm,height=3.4cm]{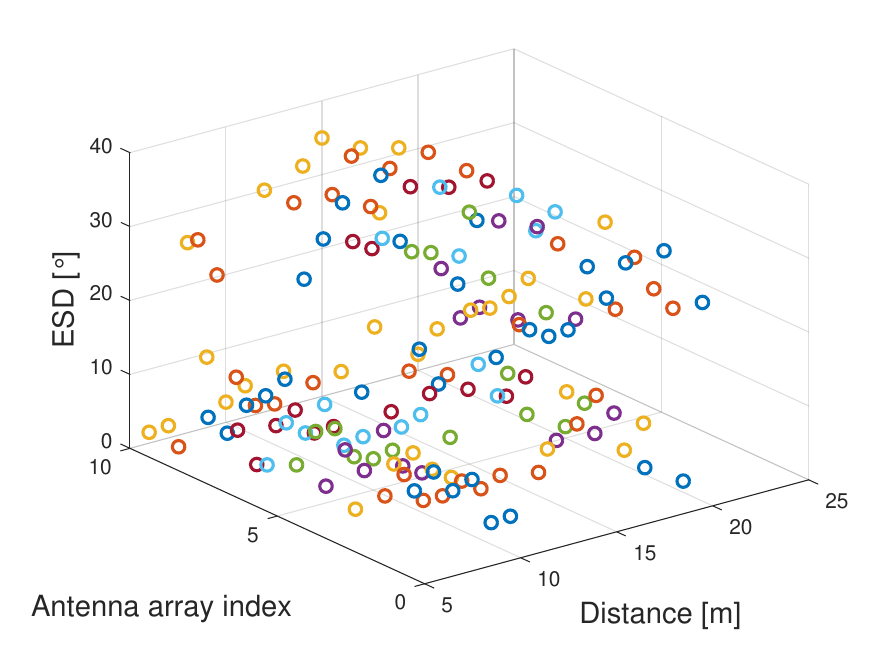} \hspace{-4mm}
		& \includegraphics[width=4.2cm,height=3.4cm]{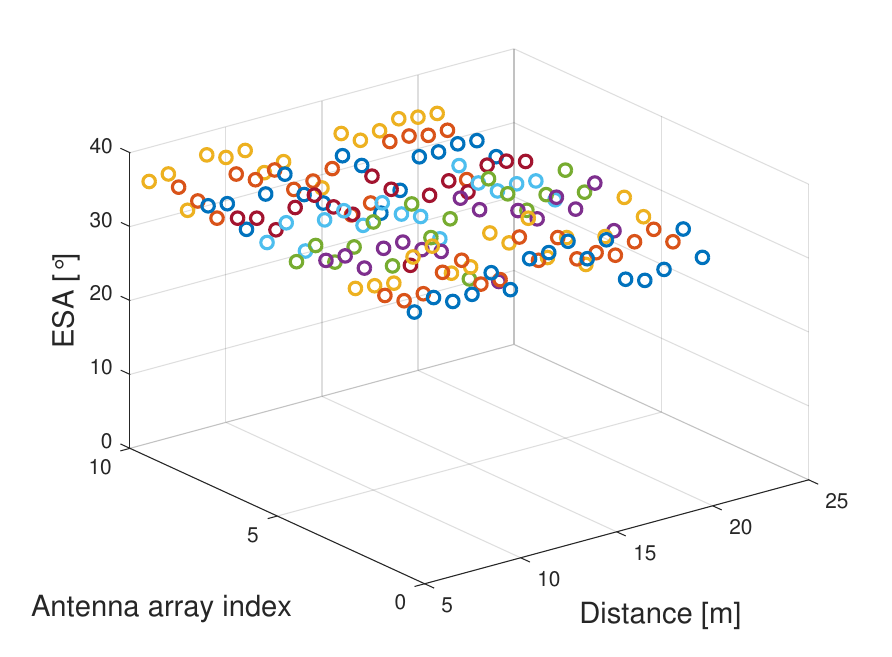}\\
		{\footnotesize\sf (a)} &	{\footnotesize\sf (b)} \\
		
	\end{tabular}
	\begin{tabular}{cc}
		\includegraphics[width=4.2cm,height=3.4cm]{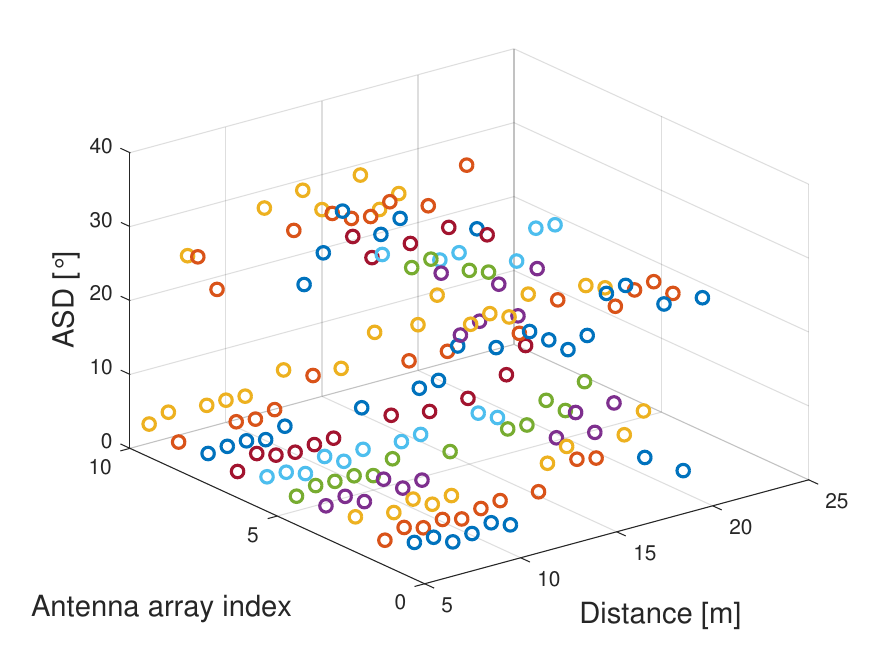} \hspace{-4mm}
		& \includegraphics[width=4.2cm,height=3.4cm]{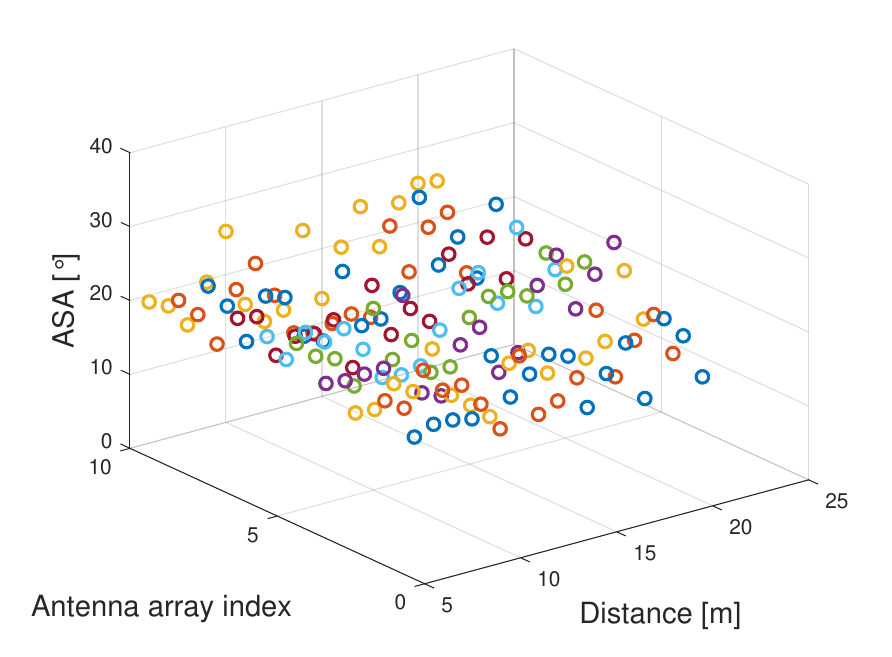}\\
		{\footnotesize\sf (c)} &	{\footnotesize\sf (d)} \\
	\end{tabular}
	\caption{ RMS angular spread in near-field (Route 1 in Case A). (a) ESD. (b) ESA. (c) ASD. (d) ASA.}
	\label{fig:Figure13}
\end{figure}

\par As shown in Fig. 13, with the changes of Rx-Tx distance and antenna array elements, the angular spread of the arrival angle does not change significantly, especially the ESA changes very little. The main reason is that the angular spread is not significant due to the small size of ODA used at the receiving end, while the obvious non-stationary change of angular spread can be seen at the transmitting end. The results show that the receiving end is usually a small user antenna in the actual communication system, so the non-stationary characteristics of the receiving end can be ignored when modeling the spatial non-stationary characteristics of the near-field.

\begin{figure}[htbp]
	\xdef\xfigwd{\columnwidth}
	\setlength{\abovecaptionskip}{0.1 cm}
	\centering
	\begin{tabular}{c}
		\includegraphics[width=0.45\textwidth]{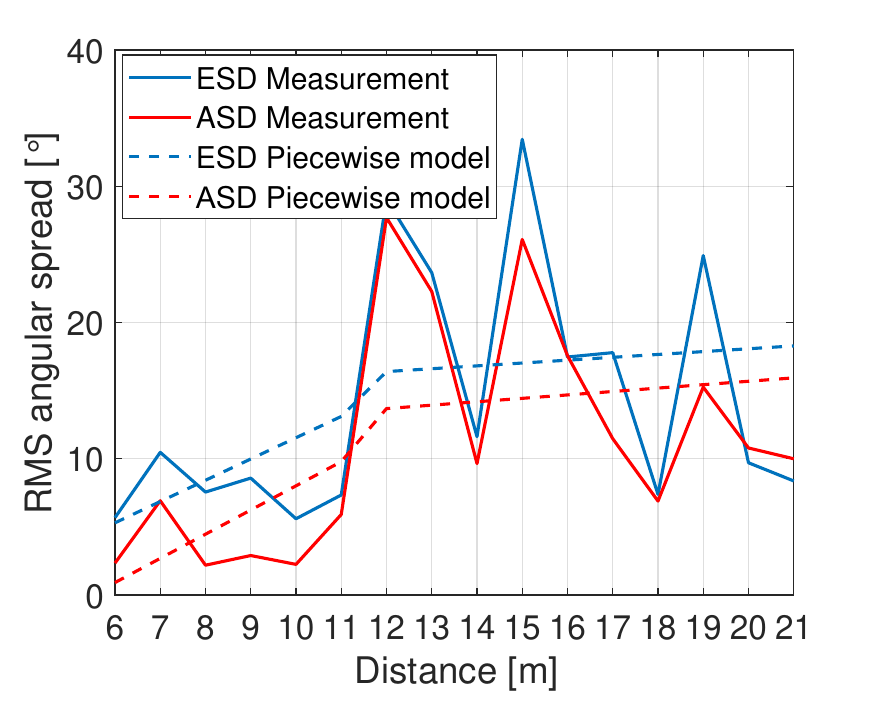}\\
		{\footnotesize\sf (a)} \\[3mm]
		\includegraphics[width=0.45\textwidth]{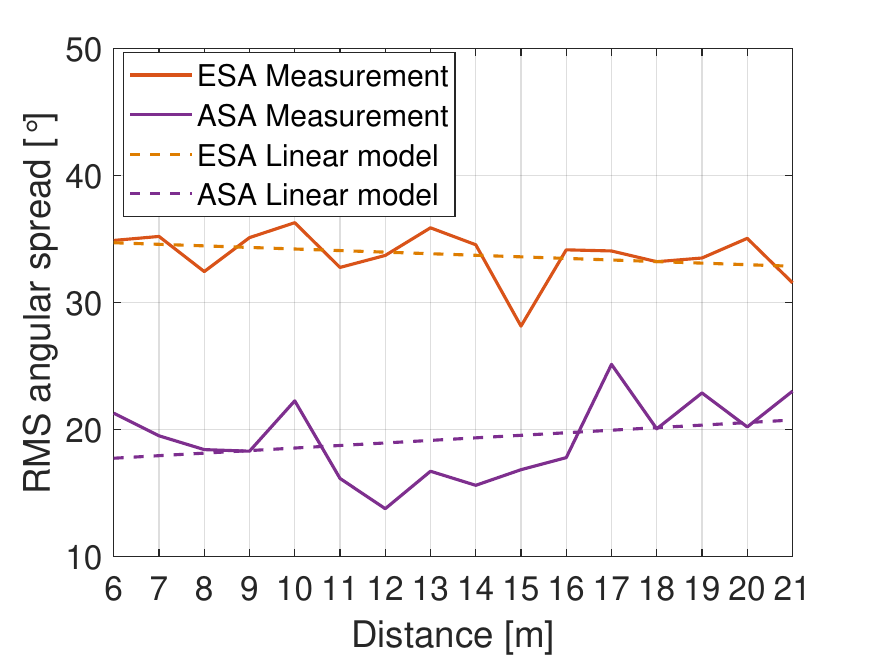}\\
		{\footnotesize\sf (b)} \\
	\end{tabular}
	\caption{The mean value of angular spread from near-field to far-field (Case B). (a) ESD and ASD. (b) ESA and ASA.}
	\label{fig:Figure14}
\end{figure}

\par From near-field to far-field, the piecewise model can be represented as

\begin{equation}
	AS = \left\{
	\begin{aligned} 
		A_1 + B_1 *d, \quad d\leq d_{BP}  \\
		AS_{BP} + B_2 *d, \quad d > d_{BP}  \\
	\end{aligned}
	\right\}   + N(0,\sigma^2)
\end{equation} 
\\ where
\emph{$A_1$} and \emph{$B_1$} are the intercept and slope before BP, respectively. \emph{$AS_{BP}$} is the delay spread at the distance of BP. \emph{$B_2$} is the slope after BP.

\begin{table}[htbp]
	\centering
	\caption{ The mean value of angular spread from near-field to far-field.}
	\setlength{\tabcolsep}{0.1 mm}
	\label{my-label}
	\renewcommand{\arraystretch}{1.5}
	\setlength{\tabcolsep}{3.5 mm}
	\begin{tabular}{c|c|c|c|c}
		\hline \hline
		Parameter    & Model &  Coefficient & $\sigma$ & $d_{BP}[m]$ \\ \hline
		ESD [$^\circ$]    &  Piecewise & \makecell{$ A_1= -4.07$, \\ $B_1= 1.56$ \\ $B_2= 0.21 $} &  2.71 & 11 \\   \hline		
		ESA [$^\circ$]    &  Linear &   \makecell{$ A_1= 35.45$, \\ $B_1= -0.12$}  & 1.94  & - \\  \hline	
		ASD [$^\circ$]   &  Piecewise &    \makecell{$ A_1= -9,74$, \\ $B_1= 1.78$ \\ $B_2= 0.25$} &  1.98 & 11 \\   \hline	
		ASA [$^\circ$]   &  Linear &   \makecell{$ A_1= 16.52$, \\ $B_1= 0.20$} & 3.07    & -\\   \hline
		\hline
	\end{tabular}
\end{table}

\par As shown in Fig. 14 and Table \uppercase\expandafter{\romannumeral6}, with the increase of Rx-Tx distance, the angular spread at the receiving end follows a linear distribution, and the slope of the change is very small (the absolute value is less than 0.3), while obvious angular expansion changes and even obvious mutations can be seen at the transmitting end, and the fluctuation range is greater than 20$^\circ$. The results show that the near field of the terminal can not be considered in the practical communication system. When the array size of the base station on the line link is large and the client (antenna size is small) is in the near-field range, the transmitting end needs to consider the near-field effect, such as using spherical wave for channel modeling.

\par In order to investigate the fluctuation degree of channel characteristics on the array, the standard deviation of angular spread on different array elements is calculated.

\begin{equation}
\begin{split}
y & = A \cdot n^2 + B \cdot n + C + N(0,\sigma^2),
\end{split} 
\end{equation}
where A, B and C are polynomial model coefficients, and $\sigma$ is the fit standard deviation.

\begin{figure}[htbp]
	\xdef\xfigwd{\columnwidth}
	\setlength{\abovecaptionskip}{0.1 cm}
	\centering
	\begin{tabular}{c}
		\includegraphics[width=0.45\textwidth]{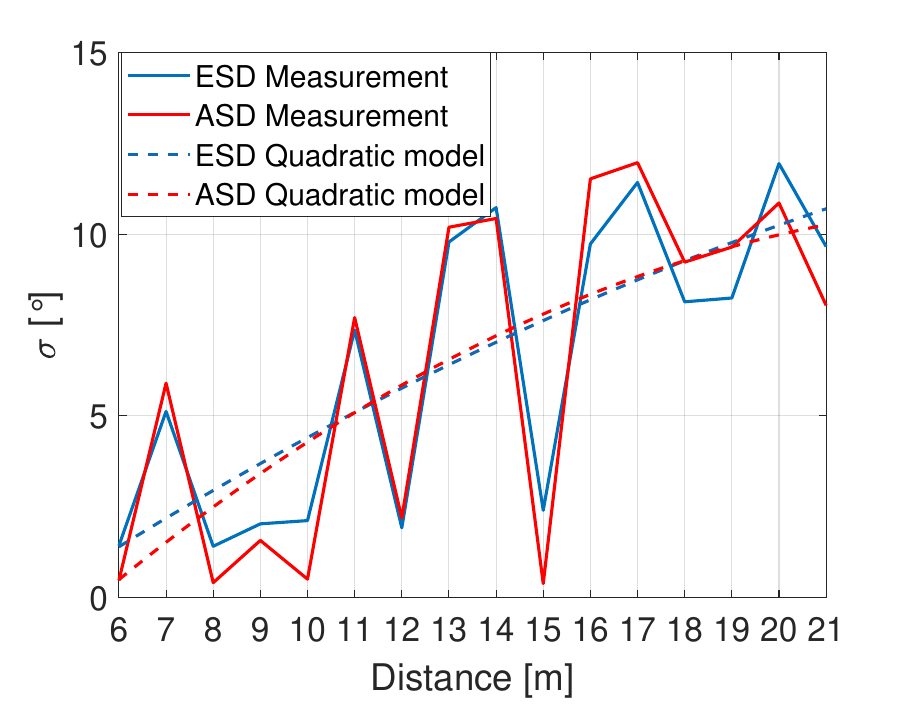}\\
		{\footnotesize\sf (a)} \\[3mm]
		\includegraphics[width=0.45\textwidth]{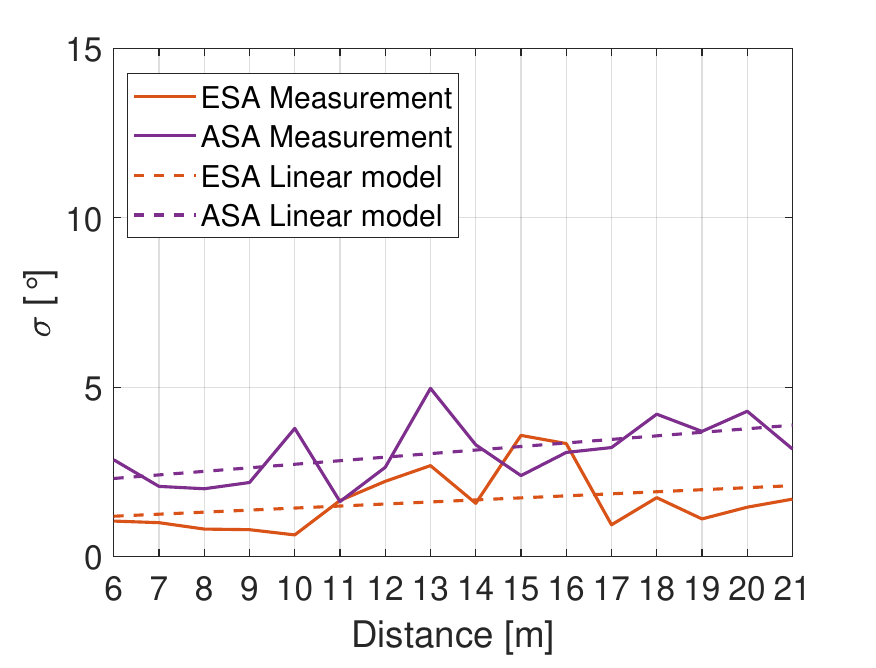}\\
		{\footnotesize\sf (b)} \\
	\end{tabular}
	\caption{The standard deviation of angular spread from near-field to far-field (Case B). (a) ESD and ASD. (b) ESA and ASA.}
	\label{fig:Figure15}
\end{figure}

\begin{table}[htbp]
	\centering
	\caption{The standard deviation of angular spread from near-field to far-field }
	\setlength{\tabcolsep}{0.1 mm}
	\label{my-label}
	\renewcommand{\arraystretch}{1.5}
	\setlength{\tabcolsep}{5 mm}
	\begin{tabular}{c|c|c|c|c}
		\hline \hline
		Parameter  &A& B& C & $\sigma$ \\ \hline
		ESD [$^\circ$]  & -0.012  & 0.94   & -3.84 & 2.89             \\ \hline
		ESA [$^\circ$]  &-   & 0.060 & 0.82  & 0.88       \\  \hline
		ASD [$^\circ$]  &-0.027   & 1.37     &-6.80 &3.57          \\  \hline
		ASA [$^\circ$]  &-  & 0.09  & 1.67 & 0.82               \\  \hline
		\hline
	\end{tabular}
\end{table}

\par As shown in Fig. 15 and Table \uppercase\expandafter{\romannumeral7}, the volatility of angular spread away from the angle is evident. With the increase of the distance of Rx-Tx, the standard deviation of the angular spread follows the quadratic distribution, and the further the distance, the fluctuation becomes slower. However, the fluctuation of the arrival angle is weak, the standard deviation follows a linear distribution, and the slope is small ($<0.1$).

\section{Conclusion}

\par In this paper, the channel measurements and characterization have been investigated for NF-FF communication. Based on TDM-MIMO mode, we designed the massive MIMO channel measurement platform. The new channel characteristics are validated based on the indoor near-field channel measurements, including delay domain and spatial domain. In the array domain, the power conforms to quadratic polynomial model distribution, and the angles of LOS path follow the linear distribution. In addition, the channel characteristics are investigated and modeled from the near-field to far-field. The spatial non-stationarity considering antenna array and transmit-receive distance are investigated and modeled the temporal-spatial channel characteristics including path loss, delay spread, and angular spread. Compared with the angular spread at the receiving end, the spatial non-stationary characteristics of angular spread at the transmitting end are more important in modeling.

\par For future work, it is necessary to further characterize the XL-MIMO channel model for 3GPP. The large and small scale channel characteristic parameters are analyzed with more measured data comprehensively. In addition, the abundant channel measurements will be conducted to build a comprehensive database of XL-MIMO channels, and generalized statistical distribution and models of channel parameters will be established.

\ifCLASSOPTIONcaptionsoff
\newpage
\fi

\end{document}